\documentstyle[psfig]{mn} 

\ifnfsstwo

\fi 

\ifnfssone 
  \newmathalphabet{\mathit} 
    \addtoversion{normal}{\mathit}{cmr}{m}{it} 
    \addtoversion{bold}{\mathit}{cmr}{bx}{it}

\fi 
 
\ifoldfss

\fi 
 
\loadboldmathitalic 
\loadboldgreek 
 
\title{The Canada-France Redshift Survey XII: 
Nature of emission-line field galaxy population up to $z=0.3$}
\author[L. Tresse {\it et al.}]  {L. Tresse$^{1,2,\dagger}$,
C. Rola$^{1,2,5}$, F. Hammer$^{2,\dagger}$, G. Stasi\'nska$^{2}$,
O. Le F\`evre$^{2,\dagger}$, \and S. J. Lilly$^{3,\dagger}$,
D. Crampton$^{4,\dagger}$ \\ $^{1}$Institute of Astronomy, Madingley
Rd., Cambridge CB3 OHA, England. \\ $^{2}$D.A.E.C., Observatoire de
Paris-Meudon, 92195 Meudon Principal Cedex, France.\\ 
$^{3}$Dept. Astronomy, Univ. of Toronto, 60 St George St., Toronto, M5S
1A7, Canada. \\ $^{4}$Dominion Astrophysical Observatory, 5071
W. Saanich Rd., R.R. 5, Victoria, B.C., V8X 4M6, Canada.\\ $^{5}$Royal
Greenwich Observatory, Madingley Rd., Cambridge CB3 OEZ, England.}
\date{Accepted 1996 February 20. Received 1995 December 27} \pubyear{1996}
 
\begin{document} 
\maketitle 
 
\begin{abstract} 
We present a spectroscopic study of the $138$ field galaxies to a
redshift $z=0.3$ from the $I$-selected Canada-France Redshift Survey.
$117$ (85\%) spectra exhibit at least $H\alpha$ in emission, the
remaining $21$ (15\%) are purely absorption-line spectra.  We focus
our analysis on spectra with H$\alpha$ and H$\beta$ in emission,
accounting for about half of this low-$z$ sample, which we classify
using emission-line ratio diagrams. Using photoionization models, we
determine the extreme boundaries of H II galaxies in these diagnostic
diagrams, and demonstrate that the emission-line ratios of a
significant fraction of galaxies require harder photoionization
sources than massive O stars.  We find that about $17 \%$ of the field
galaxies have emission-line ratios consistent with active galaxies,
{\it e.g.}  Seyfert 2 or LINERs.  After correcting for stellar
absorption under the Balmer lines, we conclude that the fraction of
such galaxies is at least $8\%$ of the field galaxy population at
$z\leq 0.3$. 
\end{abstract} 
 
 
\section{Introduction} 

Studying emission-line spectra leads to a better under\-standing of
the nature of the galaxies in deep surveys; line ratio diagrams can
separate narrow emission-line galaxies such as H II galaxies and
active galaxies.  In this paper, we adopt the terminology of ``H II
galaxies'' for galaxies with line ratios that can be explained by OB
stars as the main io\-ni\-za\-tion source of the nebula
gas, as for example starburst ga\-la\-xies (SBG), blue
compact galaxies (BCG), H II region-like galaxies.  Galaxies with line
ratios requiring harder io\-ni\-za\-tion sources than hot main sequence
stars, are gathered together under the name ``active galaxies'', as
Seyfert 2 and Low Ionization Nuclear Emitting Regions (LINER)
gala\-xies.  By observing several emission lines and considering the
appropriate line ratios, a narrow emission-line object can be
classified re\-lia\-bly (see {\it e.g.}  Veilleux \& Osterbrock 1987,
hereafter VO).
\footnotetext[2]{Visiting observer with the 
Canada-France-Hawaii Telescope, operated by the NRC of Canada, the CNRS 
of France and the University of Hawaii.} 

The wide spectral range of the $I$-selected Canada-France Redshift
Survey ($4500-8500\ \AA$) offers significant advantages for a
systematic investigation of the emission-line ratios of the field
galaxies at low redshifts. The survey is described in a series of
earlier papers (Lilly {\it et al.} 1995a (CFRS-I), Le F\`evre {\it et
al.} 1995a (CFRS-II), Lilly {\it et al.} 1995b (CFRS-III), Hammer {\it
et al.} 1995a (CFRS-IV) and Crampton {\it et al.} 1995a (CFRS-V)).  An
analysis of a few low-$z$ emission-line galaxies already showed the
presence of very strong emission-line objects from a survey
preliminary to the CFRS (Tresse {\it et al.} 1993).  At $z <0.3$,
forbidden emission lines such as [O II]$\lambda 3727$, [O III]$\lambda
4959$, [O III]$\lambda 5007$, [S II]$\lambda\lambda 6717,6731$
(hereafter [S II]$\lambda6725$) can be measured, and their ratio to
the appropriate Balmer line strength allows a classification according
to the main ionization source responsible for the emission lines. 

Our study can shed new light on the nature of the po\-pulation of faint
blue galaxies at low redshifts. Indeed, an excess of these galaxies
relative to local counts is seen in the CFRS luminosity function
(Lilly {\it et al.} 1995c (see \mbox{CFRS-VI})) for absolute magnitudes
$M(B)> -20$ mag$^{1}$.\footnotetext{$^1$Absolute magnitudes are
expressed assuming $H_0 = 50$ km s$^{-1}$ Mpc$^{-1}$ and $q_0=0.5$
throughout the paper.} A similar excess has been found previously from
$B$-selected deep surveys (Broadhurst {\it et al.} 1988, $b_{J}=21.5$;
Colless {\it et al.} 1990, $b_{J}=22.5$).  These surveys do not cover
all the optical wavelength range, hence emission-line studies have
been res\-tric\-ted mainly to the single [O II]$\lambda3727$ line. Our
systematic study of line ratios is a major step forward from previous
analyses of these galaxies, yielding a better understanding of the
evolution of this population with cosmic epoch.

The plan of this paper is as follows. In Section 2, we review some key
points concerning the CFRS observations and selection criteria which
are important for our analysis, and we describe the classification of
all the CFRS spectra up to $z=0.3$.  In Section 3 we focus on the
emission-line galaxies, describing our line measurements. In Section
4, we use two line ratio diagrams to classify these galaxies. In
Section 5, we construct photoionization models to determine the upper
limit of the H II galaxy locus in these diagnostic diagrams, and we
compare our data to this limit. In Section~6, we assess the effects of
the stellar absorption underlying the Balmer lines. Section 7 presents
a statistical analysis of our sample.  In Section 8, we investigate the
photometric characteristics of our low-$z$ sample. Section 9 discusses
our results. 

\section{Description of the sample at low redshifts}

The CFRS sample consists of $730$ $I$-band selected galaxies, of which
591 have reliable redshifts measured between $z\simeq0$ and $z=1.3$,
with a median of $\langle z \rangle=0.56$. For our analysis, we
consider the 138 spectra at $z\leq 0.3$ ($23\%$).  In this section, we
briefly summarize some key points of the survey, which are important
for our investigation, and we describe the spectral classification 
of our low-$z$ sample. 

\subsection{Characteristics of the sample}  

We have chosen to analyse only those CFRS spectra with $z\leq 0.3$,
because the H$\alpha$($\lambda 6563$) Balmer line can be seen up to
this redshift within the CFRS spectral range $(4500-8500\ \AA)$.  Using
the $H\alpha/H\beta$ emission-line ratio enables us to correct the
spectral lines for the extinction due to the interstellar dust.

From $I$-band images, the galaxies at $z<0.3$ are selected by light
from the old stellar population (light emitted above the break at
$4000\ \AA$ in the rest-frame reference), rather than on their young
stellar population. At low redshifts, a $I$-selected sample is less
sensitive to recent changes in the global star formation rate of
galaxies of various kinds, and thus is less dependent on galaxies
undergoing strong star formation than a $B$-selected sample.  The wide
$I$ band ($2000\ \AA$), centered at $8320\ \AA$, does not favour
galaxies with strong emission lines, since the equivalent width of
H$\alpha$, EW(H$\alpha$), of our emission-line galaxies is usually
less than $100\ \AA$.  The CFRS galaxies have been selected without
regard to morphology, environment or surface brightness. The CFRS
sample is limited only by $I$-isophotal apparent magnitude in the
range $17.50 \le I_{AB} \le 22.50$ mag ($I_{AB}=I+0.48$). In the
survey, this magnitude is very close to the $I$-total ap\-pa\-rent
magnitude, and thus introduces little bias in favour of compact
sources (see CFRS-I).  $85\%$ of the CFRS sample is spectroscopically
identified (see CFRS-II), and statistical arguments can be made
(CFRS-V) concerning the identifications of at least half of the
remaining galaxies.  We have estimated that the latter should have the
same redshift distribution as the identified ones, and thus the sample
is effectively $\sim 92\%$ complete.  The remaining $8\%$ likely
consists mostly of high redshift galaxies ($z>0.8$), and a few low
redshift galaxies ($z<0.1$). Below $z=0.1$, the $4000\ \AA \times
(1+z)$ break feature cannot be observed within our spectral range,
hence the redshift measurement for galaxies with very weak absorption
features becomes more difficult.

The low-$z$ CFRS sample does not favour particular kinds of galaxies, and
thus differs from emission-line samples such as those studied by
Boroson {\it et al.}  (1993) (sensitive to objects with strong
H$\alpha$ emission), Oey \& Kennicutt (1993) (H II regions in
early-type spiral galaxies), Salzer {\it et al.}  (1989)
(emission-line galaxy sample).
  
From these overall key points, the CFRS is expected to provide a
sample at $z\leq 0.3$ which is a fair representation of the field
galaxies with absolute magnitudes $-22 <M(B_{AB})<-14$ mag,
($B_{AB}=B-0.17$) (see \mbox{CFRS-VI}).

The CFRS sample consists of five fields (00h, 03h, 10h, 14h and 22h)
of area $10 \times 10$ arcmins, chosen at high galactic latitude
($|b_{II}|\ge 45^{o}$) (see CFRS-I).  The spectra are flux ca\-li\-bra\-ted with
spectrophotometric standard stars (see \mbox{CFRS-II}).

\subsection{Spectral classification} 

The $138$ spectra are classified using only the
H$\alpha$($\lambda6563$) and H$\beta$($\lambda4861$) Balmer lines
regardless of other features, as follows (see Figure~\ref{spectres}).
Class (A) consists of spectra with H$\alpha$ and H$\beta$ in
emission. Class (B) consists of spectra with H$\alpha$ in emission and
H$\beta$ in absorption. Class (C) consists of spectra with H$\alpha$
and H$\beta$ in absorption.  Absorption and emission line features
reflect different physical phenomena.  Balmer absorption lines are
produced in stellar atmospheres, whereas Balmer emission lines are
produced by hot gas sur\-roun\-ding stars.  Absorption lines are tracers
of stellar populations in galaxies.  Emission lines are tracers of
violent phenomena such as star formation, supernovae, or even harder
radiation sources such as active nuclei.

Spectra of class (A) show a blue continuum.  They are strong
emission-line galaxies: the rest-frame EW(H$\alpha$) lies in the range
$20 < EW(H\alpha) <100\ \AA$, and the spectra exhibit forbidden lines
such as [O III]$\lambda4959$, [O III]$\lambda5007$,
[O~II]$\lambda3727$ and [S II]$\lambda 6725$.  Spectra of class (B)
show a red continuum, and they exhibit moderate to strong H$\alpha$
emission, which indicates galaxies with a range of star formation
rates.  Several absorption features such as H$\beta$, the 4000 $\AA$
break, the G-band ($\lambda4304$) and MgI ($\lambda5183$) trace the
underlying stellar population. They look like starbursts oc\-cu\-ring in
early-type galaxies.  Spectra of class (C) show also a red continuum,
but they exhibit only absorption lines.
\begin{figure*}
\caption{\label{spectres} Examples of spectra in class (A) (top left), 
in sub-class (AB) (bottom left), in class (B) (top right) and in class
 (C) (bottom right).}
\end{figure*}
 
Lines are not resolved by the CFRS spectral resolution ($40\ \AA$),
therefore have velocity widths less than $1000$~km~s$^{-1}$; this 
characterizes our objects as narrow-line galaxies.  No spectrum with
broad Balmer emission lines typical of quasars or Seyfert 1 galaxies,
are present up to $z=0.3$. However, the presence of weak broad line
components underlying narrow emission Balmer lines could not be
detected with our low resolution.

Some spectra with H$\alpha$ in emission do not exhibit either H$\beta$
in emission like those in class (A), or in absorption like those in
class (B).  Such spectra are included in class (A) because they show
spectral features typical of strong emission-line spectra such as [O
II]$\lambda 3727$, [O III]$\lambda 4959$, [O~III]$\lambda 5007$, [S
II]$\lambda6725$. In these cases, stellar absorption at H$\beta$ must
be significant, and thus hides the H$\beta$ emission. The ($40\ \AA$)
resolution cannot resolve these opposite effects, and thus no accurate
measurement can be done.  We investigate this crucial point in more
detail below.  We classify these spectra in sub-class (AB) (see
Figure~\ref{spectres}).

The number of spectra by spectral class and CFRS field are given in
Table~\ref{class}.  $85\%$ of all galaxies up to $z=0.3$ are
emission-line galaxies (classes A and B); this is a high fraction,
especially considering they are low-$z$, $I$-selected faint field
galaxies. Nevertheless, there is no published fraction of
emission-line galaxies in the nearby Universe to compare reliably
with.  In their $B$-selected sample of faint field galaxies,
Glazebrook {\it et al.} (1995) find $\sim 73\%$ (16/22) of
emission-line galaxies at $z<0.3$. Although lower than our result,
this is comparable; spectra without [O II]$\lambda3727$ in class (B)
would have been accounted for absorption line spectra in their
classification since at $z>0.2$ H$\alpha$ is usually outside their
spectral range.  The single class (A) represent $53\%$, and thus
indicates that about half of the field galaxy population undergoes
violent phenomena at $z\leq 0.3$.  An emission-line spectrum exhibits
at least $H\alpha$ in emission, the most reliable quantitative tracer
of massive star formation ({\it e. g.} Kennicutt 1992) if only star
formation is occuring.  However, in galaxies with an active nucleus,
H$\alpha$ is the result of both star formation and nuclear
activity. Therefore to derive star formation rates, one must be
careful to separate these two contributions.
\begin{table}
\begin{center}
\begin{tabular}{|c|cc|c|c||c|}
\hline 
Fields &  \multicolumn{5}{c|}{Classes} \\
   \cline {2-6}
\hfill \\
&  A & (AB) & B & C & Total\\
\hline 
00h & 12 & (5) & 8 & 4 & 24 \\
03h & 20 & (3) & 15 & 2 & 37 \\
10h & 11 & (1) & 9 & 3 & 23 \\
14h & 21 & (4) & 6 & 5 & 32 \\
22h & 10 & (5) & 5 & 7 & 22 \\
\hline
Total & 74 & (18) & 43 & 21 & 138 \\
\hline
\hline
Percentages & 53.6 \% & (13 \%) & 31.1 \% & 15.3 \% & 100 \% \\
\hline
\end{tabular} 
\end{center}
\caption{\label{class}Distribution of the $138$ spectra at $z\leq 0.3$   
  from the CFRS sample by spectral classes as described in Section 2.}
\end{table}

In this paper, we investigate in detail emission-line galaxies of
class (A), $74$ in total, using emission-line ratio dia\-grams.  Details
of the spectra which are analysed, are listed in Table~\ref{tableflux}.
\hfill \\
\hfill \\

\section{Data Analysis} 
%
\tiny
\begin{table*} 
\flushleft
\begin{tabular}{|clrrrrrl|}
\hline 
& & & & & & & \\ 
Id. & A$_{V}$ & 
H$\beta$\hspace*{0.775cm} &  H$\alpha$\hspace*{0.775cm} & [OII]$\lambda3727$  & [OIII]$\lambda5007$  & 
[SII]$\lambda6725$ & Type \\
(1)& (2)& (3)\hspace*{0.775cm}  & (4)\hspace*{0.775cm}& (5)\hspace*{0.775cm}& (6)\hspace*{0.775cm}& (7)\hspace*{0.775cm}& (8) \\   
\hline
& & & & & & & \\ 
00.0121 & 2.4840 & 11.340 $\pm$  0.665 &  75.731 $\pm$ 2.243 & 46.936 $\pm$ 3.817  & 
       95.676 $\pm$ 1.116 &  $-$\hspace*{0.775cm}  & Act. \\ 
00.0159 & 4.7384 &  1.000\hspace*{0.6cm}  & 14.419 $\pm$ 0.849  & $-$\hspace*{0.775cm} & 
     12.356 $\pm$ 0.802 &  4.082 $\pm$ 0.654  & Act. \\
00.0343 & 3.6510 & 0.516\hspace*{0.6cm}        & 5.135 $\pm$ 0.790 & 6.203 $\pm$ 1.353  & 
    1.725 $\pm$ 0.469 &  2.394 $\pm$ 0.551 & Act. \\
00.0351 & 0.0000 & 5.527 $\pm$ 1.075 &  7.097 $\pm$ 1.240 &  6.379 $\pm$ 1.437  &   
   6.403 $\pm$ 1.132 &  0.700\hspace*{0.6cm} & H II   \\ 
00.1013 & 3.6684 & 3.250 $\pm$ 0.300 & 32.521 $\pm$ 3.258 & 21.026 $\pm$ 3.711 & 
   12.713 $\pm$ 1.771 & 17.187 $\pm$ 2.343 & Act.\\
00.1057 & 3.1505 & 0.508\hspace*{0.6cm}       & 4.259 $\pm$ 1.031 & 7.159 $\pm$ 0.770  & 
   6.500 $\pm$ 0.523 &  2.480 $\pm$ 0.795 & Act. \\
00.1203 & 2.7999 & 1.290 $\pm$ 0.300 &  9.596 $\pm$ 0.729 &  2.804 $\pm$ 0.870 &  
    4.319 $\pm$ 0.594 &  5.121 $\pm$ 0.612 & Act.\\ 
00.1444 & 2.0011 & 1.028 $\pm$ 0.173 & 5.823 $\pm$ 0.726 & 3.071 $\pm$ 0.864  &  
       8.508 $\pm$ 0.258 &  $-$\hspace*{0.775cm}  & Act.   \\ 
00.1649 & 1.2973 & 1.940 $\pm$ 0.250 & 8.640 $\pm$ 0.610 & 17.052 $\pm$ 2.915  & 
    8.833 $\pm$ 1.308 &  3.500\hspace*{0.6cm} & Act. \\
00.1903 & 0.7370 & 0.700\hspace*{0.6cm} & 2.575 $\pm$ 0.630 & 3.000\hspace*{0.6cm}  & 
    2.903 $\pm$ 0.499 &  0.760\hspace*{0.6cm}& Inter.   \\
03.0003 & 0.4195 & 1.315 $\pm$ 0.320 &  4.340 $\pm$ 0.345 & 1.299 $\pm$ 0.716  &  
     9.339 $\pm$ 0.700 &  0.727\hspace*{0.6cm} & Inter. \\
03.0096 & 1.3063 & 1.783 $\pm$ 0.431 &  7.967 $\pm$ 1.356 &  3.258 $\pm$ 0.751 &  
    10.782 $\pm$ 0.665  &  1.248 $\pm$ 0.530 & H II  \\
03.0165 & 0.9098 & 26.209 $\pm$ 1.051 & 102.262 $\pm$ 2.993 & 132.039 $\pm$ 6.056 & 
    113.883 $\pm$ 1.470 & 10.386 $\pm$ 2.103  & H II \\
03.0241 & 1.8488 &  0.930\hspace*{0.6cm}  &  5.000 $\pm$ 1.500 &   4.003 $\pm$ 0.810  &  
       4.003 $\pm$ 0.545 &  $-$\hspace*{0.775cm}  & Act.  \\ 
03.0315 & 7.2036 & 1.070 $\pm$ 0.700 &  35.795 $\pm$ 1.241 & 24.531 $\pm$ 0.714 & 
    9.591 $\pm$ 0.672 &  9.094 $\pm$ 0.915  & Act. \\
03.0332 & 1.7032 &   1.601\hspace*{0.6cm} & 8.190 $\pm$ 0.600 &  1.782 $\pm$ 0.600 
 &  2.444 $\pm$ 0.300  &  2.710 $\pm$ 0.606 & H II  \\ 
03.0443 & 1.4470 & 134.902 $\pm$ 2.441 & 632.303 $\pm$ 8.645 & $-$\hspace*{0.775cm}      & 
   740.424 $\pm$ 4.004 & 76.834 $\pm$ 4.508 & H II \\ 
03.0562 & 1.4601 & 22.493 $\pm$ 0.883 & 105.904 $\pm$ 1.439 & $-$\hspace*{0.775cm}     &  
   80.109 $\pm$ 1.028 & 25.001 $\pm$ 1.011 & H II  \\
03.0605 & 1.8952 & 3.194 $\pm$ 0.506 & 17.445 $\pm$ 0.993 & 9.687 $\pm$ 1.166  & 
     20.372 $\pm$ 0.767 &  3.263 $\pm$ 0.716 & H II  \\
03.0982 & 2.0389 & 2.809 $\pm$ 0.396 & 16.117 $\pm$ 0.596 & 9.792 $\pm$ 1.332  & 
   12.217 $\pm$ 0.446 &  3.818 $\pm$ 0.431 & H II \\ 
03.0996 & 1.4580 & 1.050\hspace*{0.6cm}  & 4.940 $\pm$ 0.500 &  3.278 $\pm$ 1.000  & 
   4.084 $\pm$ 0.400 &  1.850 $\pm$ 0.600  & Act.  \\ 
03.1051 & 4.9073 & 1.730\hspace*{0.6cm}  &  26.425 $\pm$ 0.967 & $-$\hspace*{0.775cm} & 
    16.952 $\pm$ 1.370 &  5.928 $\pm$ 0.694  & Act. \\ 
03.1097 & 1.2897 & 1.405 $\pm$ 0.400 & 6.241 $\pm$ 0.597 &  $-$\hspace*{0.775cm}        & 
   3.630 $\pm$ 0.546 &  1.349 $\pm$ 0.400  & H II    \\ 
03.1179 & 3.3531 &   1.500 $\pm$ 0.300 &  13.478 $\pm$ 0.991 & 0.878\hspace*{0.6cm} &  
       6.279 $\pm$ 0.637 &  $-$\hspace*{0.775cm}   & H II   \\ 
03.1343 & 1.3014 & 3.092 $\pm$ 0.500 & 13.790 $\pm$ 0.785 & 10.500 $\pm$ 2.000 & 
   8.418 $\pm$ 0.500 &  4.474 $\pm$ 0.983  & H II \\ 
10.0523 & 0.0000 & 7.664 $\pm$ 1.093 & 21.500 $\pm$ 2.248 &  68.661 $\pm$ 5.556 & 
    32.633 $\pm$ 1.000 & 21.195 $\pm$ 2.122 & Act.\\
10.1178 & 0.7569 & 2.813 $\pm$ 0.741 & 10.418 $\pm$ 0.762 & 30.912 $\pm$ 3.765  & 
   11.193 $\pm$ 0.947 &  3.008 $\pm$ 0.592  & Inter.\\ 
10.1643 & 5.8490 & 2.000\hspace*{0.6cm}   & 42.134 $\pm$ 1.286  & $-$\hspace*{0.775cm} & 
     12.134 $\pm$ 1.456 & 11.970 $\pm$ 0.929  & Inter. \\ 
10.1650 & 1.4862 & 934.571 $\pm$ 8.070 &  196.739 $\pm$ 3.141 & $-$\hspace*{0.775cm}   & 
   910.474 $\pm$ 4.820 & 95.232 $\pm$ 4.483  & H II  \\
10.1653 & 1.0553 & 7.302 $\pm$ 1.200  & 29.940 $\pm$ 1.348 & 13.388 $\pm$ 2.296  & 
   23.829 $\pm$ 1.409 &  9.120 $\pm$ 0.999  & H II \\   
10.1889 & 3.7950 & 3.110 $\pm$ 0.400 & 32.494 $\pm$ 1.961 & 12.130 $\pm$ 2.564  &  
    9.249 $\pm$ 1.383 &  6.341 $\pm$ 1.227 & H II  \\ 
10.1899 & 0.8123 & 32.867 $\pm$ 2.165 & 124.042 $\pm$ 2.189 &  $-$\hspace*{0.775cm}    & 
  165.288 $\pm$ 2.659 & 35.450 $\pm$ 1.835  & Inter.\\
14.0025 & 2.6973 & 7.184 $\pm$ 1.534 &  51.599 $\pm$ 2.698 & 32.913 $\pm$ 6.854  & 
   22.200 $\pm$ 1.369 & 20.403 $\pm$ 1.973  & Inter. \\ 
14.0070 & 1.2116 & 7.323 $\pm$ 2.000 & 31.673 $\pm$ 0.983 &  $-$\hspace*{0.775cm}   &  
   22.575 $\pm$ 2.831 &  9.213 $\pm$ 0.694  & H II  \\ 
14.0208 & 0.9112 & 2.500\hspace*{0.6cm}  & 9.759 $\pm$ 1.239 &  11.409 $\pm$ 2.290  & 
   11.954 $\pm$ 1.492  &  3.932 $\pm$ 1.000 & Act.   \\ 
14.0310 & 1.7163 & 8.189 $\pm$ 0.600 & 42.080 $\pm$ 1.000 &   23.600 $\pm$ 1.000 & 
    23.340 $\pm$ 0.700  &  8.035 $\pm$ 0.600  & H II  \\ 
14.0377 & 1.5170 & 12.100 $\pm$ 0.642 &  58.086 $\pm$ 0.849 &  1.517\hspace*{0.6cm} & 
     28.647 $\pm$ 1.989 & 34.091 $\pm$ 0.720 & Act. \\ 
14.0528 & 1.5137 & 13.964 $\pm$ 0.930 &  66.957 $\pm$ 1.369 & $-$\hspace*{0.775cm}   & 
     64.860 $\pm$ 1.273 & 12.003 $\pm$ 0.978  & H II  \\ 
14.0588 & 0.0000 & 5.462 $\pm$ 0.881 &  7.830 $\pm$ 0.630 &   $-$\hspace*{0.775cm}   & 
    19.803 $\pm$ 0.964 &  6.273 $\pm$ 0.628  & Act. \\ 
14.0621 & 3.1920 & 2.716 $\pm$ 0.512 & 23.100 $\pm$ 1.489 & 23.157 $\pm$ 2.297  &  
    12.307 $\pm$ 0.641 &  9.710 $\pm$ 1.077  & Act. \\ 
14.1039 & 2.2419 & 54.634 $\pm$ 2.434 & 335.920 $\pm$ 2.765 &  $-$\hspace*{0.775cm}      & 
    231.396 $\pm$ 2.876 & 77.910 $\pm$ 2.083 & H II  \\ 
14.1103 & 0.0000 & 120.195 $\pm$ 7.080 & 46.932 $\pm$ 2.296 & 4.220\hspace*{0.6cm} &
     327.016 $\pm$ 5.437  &  0.743\hspace*{0.6cm} & H II  \\
14.1209 & 6.0558 & 1.000\hspace*{0.6cm} &  22.608 $\pm$ 1.394 & $-$\hspace*{0.775cm} &   
     6.757 $\pm$ 0.842 &  2.028 $\pm$ 0.765 & H II  \\ 
14.1273 & 0.5670 & 6.903 $\pm$ 0.816 &  23.959 $\pm$ 1.658 & 11.623 $\pm$ 1.751  &  
   18.573 $\pm$ 0.890   &  4.832 $\pm$ 1.108 & H II \\ 
14.1376 & 0.5945 & 6.816 $\pm$ 1.589 &  23.880 $\pm$ 0.300 &  35.233 $\pm$ 1.860  & 
      9.461 $\pm$ 1.619 &  $-$\hspace*{0.775cm}   & Inter.   \\ 
14.1425 & 1.6956 & 2.916 $\pm$ 0.665 & 14.878 $\pm$ 1.020 &  6.074 $\pm$ 1.037  &  
    9.319 $\pm$ 0.918 &  1.155\hspace*{0.6cm} & H II  \\
22.0474 & 0.3030 & 9.585 $\pm$ 0.376 &  30.400 $\pm$ 0.200 &  15.621 $\pm$ 1.723  & 
   38.187 $\pm$ 0.530 &  7.526 $\pm$ 0.797   & H II  \\ 
22.0676 & 0.6017 & 7.006 $\pm$ 0.590 & 24.608 $\pm$ 0.934  & $-$\hspace*{0.775cm} & 
    22.812 $\pm$ 0.784  & 10.487 $\pm$ 0.697 & Act. \\ 
22.1013 & 1.0590 & 4.189 $\pm$ 0.442 & 17.198 $\pm$ 0.727 & 15.500 $\pm$ 1.600  &  
    8.897 $\pm$ 0.479 &  8.000 $\pm$ 0.614  & H II  \\ 
22.1082 & 0.5644 & 9.346 $\pm$ 0.977 &  32.409 $\pm$ 0.032 & 20.310 $\pm$ 2.078  & 
      41.137 $\pm$ 1.445 &  $-$\hspace*{0.775cm}   & H II    \\ 
& & & & & & & \\ 
\hline 
\end{tabular}
\normalsize
\caption{\label{tableflux} 
Spectra analysed in class (A). Meaning of the columns: (1) CFRS
identification (($00.\star\star\star\star$ are galaxies in the 00h
field, $10.\star\star\star\star$, those in the 10h field, etc., RA
and DEC are published in CFRS-II, CFRS-III and CFRS-IV), (2) reddening
in magnitude as calculated in Section 3.2, (3)-(7) fluxes in
10$^{-17}$ erg s$^{-1}\ \AA^{-1}$ with 1 $\sigma$ error respectively
for H$\beta$, H$\alpha$, [O II]$\lambda3727$, [O III]$\lambda5007$ and
[S II]$\lambda6725$ (the fluxes with no error range are the fluxes
taken equal at 1 $\sigma$ level as explained in Section 3.1), (8)
classification of galaxies derived from Figure 9, ``Act.'', ``H II''
and ``Inter.'' mean respectively active galaxies (solid points), H II
galaxies (open points), and intermediate galaxies (open points with a
cental dot).}
\end{table*} 
\normalsize

\tiny
\begin{table*}
\flushleft
\begin{tabular}{|clcrcrrrrr|}
\hline 
& & & & & & & & & \\ 
Id. & z & I$_{AB}$ & (V-I)$_{AB}$ & M(B$_{AB}$) & 
EW([OII]) & EW(H$\beta$) & EW([O III]) & EW(H$\alpha$) & EW([S II])  \\
(1) & (2) & (3) & (4) & (5) & (6)\hspace{0.5cm} & (7)\hspace*{0.4cm} & (8)\hspace*{0.6cm} & (9)\hspace*{0.4cm} & 
(10)\hspace*{0.4cm} \\   
\hline
& & & & & & & & &  \\ 
00.0121 & 0.3000 & 21.22 & 0.23 & $-$19.71 & 71\ \ \ 64\ \ \ 73 & 20\ \ \ 16\ \ \ 25 & 175\ $\,$201\ $\,$187 & 254$\ \,$217$\ \,$291 & $-$\hspace*{0.6cm} \\ 
00.0159 & 0.0866 & 20.67 & 0.39 & $-$17.33 & $-$\hspace*{0.6cm} & 2\ \ \ \ $\,$1\ \ \ \ $\,$3 & 18\ \ \ 16\ \ \ 21  & 30\ \ \ 28\ \ \ 33 & 10\ \ \ $\,$9\ \ \ 11 \\
00.0343 & 0.2474 & 22.08 & 0.35 & $-$18.27 & 48\ \ \ 40\ \ \ 51 & 2\ \ \ \ $\,$1\ \ \ \ $\,$3  & 11\ \ \ \ $\,$9\ \ \ 13 & 34\ \ \ 30\ \ \ 40 & 16\ \ \ 15\ \ \ 17  \\
00.0351 & 0.2279 & 22.02 & 0.48 & $-$18.00 & 45\ \ \ 33\ \ \ 57 &  22\ \ \ 16\ \ \ 38 & 30\ \ \ 24\ \ \ 43  & 69\ \ \ 53\ \ \ 94 & 14\ \ \ 10\ \ \ 24 \\ 
00.1013 & 0.2445 & 20.74 & 0.59 & $-$19.34 & 21\ \ \ 14\ \ \ 23 & 5\ \ \ \ $\,$4\ \ \ \ $\,$7 & 19\ \ \ 15\ \ \ 22 & 63\ \ \ 48\ \ \ 78 & 35\ \ \ 29\ \ \ 45 \\
00.1057 & 0.244  & 22.17 & 0.63 & $-$17.86 & 58\ \ \ 46\ \ \ 68 & 4\ \ \ \ $\,$3\ \ \ \ $\,$6 &  43\ \ \ 37\ \ \ 58 & 40\ \ \ 30\ \ \ 58 & 26\ \ \ 21\ \ \ 38  \\
00.1203 & 0.2467 & 21.81 & 0.67 & $-$18.20 & 15\ \ \ 13\ \ \ 17 & 6\ \ \ \ $\,$5\ \ \ \ $\,$9  & 21\ \ \ 19\ \ \ 25 & 52\ \ \ 42\ \ \ 58 & 27\ \ \ 21\ \ \ 34  \\ 
00.1444 & 0.2462 & 21.61 & 1.11 & $-$17.94 & 7\ \ \ \ \ $\,$5\ \ \ \ $\,$9 & 4\ \ \ \ $\,$3\ \ \ \ $\,$6 & 47\ \ \ 44\ \ \ 53 & 38\ \ \ 33\ \ \ 44 & $-$\hspace*{0.6cm} \\ 
00.1649 & 0.2725 & 21.35 & 0.59 & $-$19.00 & 43\ \ \ 36\ \ \ 53 &  6\ \ \ \ $\,$5\ \ \ \ $\,$9 & 22\ \ \ 18\ \ \ 27 & 60\ \ \ 48\ \ \ 64 & 15\ \ \ 10\ \ \ 19 \\
00.1903 & 0.2281 & 22.22 & 0.70 & $-$17.56 & 10\ \ \ \ $\,$7\ \ \ 17  & 5\ \ \ \ $\,$4\ \ \ \ $\,$8 & 28\ \ \ 24\ \ \ 35 & 19\ \ \ 14\ \ \ 28 & 4\ \ \ \ $\,$2\ \ \ $\,$5  \\ 
03.0003 & 0.2205 & 22.49 & 1.31 & $-$16.60 & 52\ \ \ 45\ \ \ 60 & 26\ \ \ 20\ \ \ 38 & 137\ $\,$112\ $\,$147 & 70\ \ \ 60\ \ \ 94 & 11\ \ \ \ $\,$7\ \ \ 17\\
03.0096 & 0.2173 & 22.14 & 0.93 & $-$17.24 & 27\ \ \ 20\ \ \ 36 & 10\ \ \ \ $\,$8\ \ \ 15 & 70\ \ \ 57\ \ \ 82 & 57\ \ \ 46\ \ \ 78  & 7\ \ \ \ $\,$5\ \ \ 13 \\
03.0165 & 0.177  & 20.04 & 0.40 & $-$19.50 & 68\ \ \ 60\ \ \ 71 & 16\ \ \ 14\ \ \ 18 & 60\ \ \ 56\ \ \ 64 & 75\ \ \ 65\ \ \ 86 &  8\ \ \ \ $\,$6\ \ \ 10 \\
03.0241 & 0.2896 & 22.02 & 0.59 & $-$18.48 & 41\ \ \ 30\ \ \ 55  & 4\ \ \ \ $\,$2\ \ \ \ $\,$6 & 27\ \ \ 24\ \ \ 32 & 15\ \ \ 11\ \ \ 20 & $-$\hspace*{0.6cm}\\ 
03.0315 & 0.2228 & 20.28 & 1.40 & $-$18.83 & 59\ \ \ 50\ \ \ 64 &  3\ \ \ \ $\,$2\ \ \ \ $\,$5 & 16\ \ \ 14\ \ \ 19 & 65\ \ \ 54\ \ \ 78 & 16\ \ \ 12\ \ \ 23 \\
03.0332 & 0.1877 & 21.88 & 0.53 & $-$17.63 & 14\ \ \ 11\ \ \ 15 & 7\ \ \ \ $\,$5\ \ \ \ $\,$8  & 12\ \ \ 11\ \ \ 15 & 48\ \ \ 42\ \ \ 54 & 14\ \ \ \ $\,$9\ \ \ 18 \\ 
03.0443 & 0.1178 & 19.48 & $-$0.02 & $-$19.76 & $-$\hspace*{0.6cm} & 54\ \ \ 51\ \ \ 57 & 337\ $\,$328\ $\,$373 & 376\ $\,$357\ $\,$394 & 52\ \ \ 46\ \ \ 59 \\ 
03.0562 & 0.1689 & 19.91 & 0.37 & $-$19.55 & $-$\hspace*{0.6cm} & 10\ \ \ \ $\,$8\ \ \ 12 & 45\ \ \ 43\ \ \ 49 & 84\ \ \ 77\ \ \ 92 & 24\ \ \ 18\ \ \ 29 \\
03.0605 & 0.2193 & 21.28 & 0.36 & $-$18.78 & 38\ \ \ 26\ \ \ 42 & 11\ \ \ \ $\,$8\ \ \ 14 & 55\ \ \ 50\ \ \ 62 & 84\ \ \ 67\ \ \ 99 & 13\ \ \ \ $\,$9\ \ \ 18 \\
03.0982 & 0.1952 & 21.26 & 0.68 & $-$18.15 & 56\ \ \ 48\ \ \ 60 &  8\ \ \ \  $\,$6\ \ \ \ $\,$9 & 34\ \ \ 31\ \ \ 37 & 55\ \ \ 48\ \ \ 61 & 11\ \ \ \ $\,$8\ \ \ 13 \\ 
03.0996 & 0.2193 & 22.09 & 0.63 & $-$17.67 & 85\ \ \ 52\ \ \ 90 & 4\ \ \ \ $\,$3\ \ \ \ $\,$7 & 21\ \ \ 18\ \ \ 26  & 36\ \ \ 27\ \ \ 43 & 10\ \ \ \ $\,$6\ \ \ 13 \\
03.1051 & 0.1554 & 20.77 & 0.59 & $-$18.22 & $-$\hspace*{0.6cm} &  2\ \ \ \ $\,$1\ \ \ \ $\,$3 & 17\ \ \ 16\ \ \ 18 & 38\ \ \ 36\ \ \ 41 & 12\ \ \ 10\ \ \ 13\\ 
03.1097 & 0.198  & 22.40 & 0.98 & $-$16.66 & $-$\hspace*{0.6cm} & 13\ \ \ \ 8\ \ \ 20 & 36\ \ \ 30\ \ \ 47 & 44\ \ \ 31\ \ \ 62 & 10\ \ \ \ $\,$6\ \ \ 15 \\  
03.1179 & 0.1837 & 21.45 & 0.52 & $-$18.03 & 6\ \ \ \ $\,$4\ \ \ 10 & 3\ \ \ \ $\,$2\ \ \ \ $\,$4 & 22\ \ \ 21\ \ \ 24 & 46\ \ \ 38\ \ \ 57 & $-$\hspace*{0.6cm} \\ 
03.1343 & 0.1887 & 21.69 & 0.40 & $-$17.99 & 191\ $\,$105\ $\,$202 & 11\ \ \ \ $\,$9\ \ \ 13 & 32\ \ \ 28\ \ \ 37 & 86\ \ \ 74\ \ 105 & 26\ \ \ 21\ \ \ 30 \\ 
10.0523 & 0.1949 & 20.63 & 0.49 & $-$19.02 & 56\ \ \ 49\ \ \ 62 & 9\ \ \ \ $\,$7\ \ \ 11 & 33\ \ \ 30\ \ \ 36 & 45\ \ \ 41\ \ \ 50 & 28\ \ \ 25\ \ \ 33  \\
10.1178 & 0.1971 & 21.57 & 0.61 & $-$17.95 & $-$\hspace*{0.6cm} & 13\ \ \ 11\ \ \ 15 & 47\ \ \ 44\ \ \ 52 & 66\ \ \ 60\ \ \ 79 & 19\ \ \ 15\ \ \ 23 \\ 
10.1643 & 0.2345 & 20.77 & 0.84 & $-$18.93 & $-$\hspace*{0.6cm} &  3\ \ \ \ $\,$2\ \ \ \ $\,$4 & 30\ \ \ 27\ \ \ 35  & 94\ \ \ 84\ \ 104  & 32\ \ \ 27\ \ \ 36\\ 
10.1650 & 0.0067 & 17.96 & 0.18 & $-$14.89 & $-$\hspace*{0.6cm} & 56\ \ \ 54\ \ \ 60 & 273\ $\,$256\ $\,$323 & 370\ $\,$334\ $\,$400 & 43\ \ \ 35\ \ \ 57 \\
10.1653 & 0.1955 & 21.49 & 0.51 & $-$18.14 & 87\ \ \ 57 100 &  13\ \ \ 11\ \ \ 16 & 49\ \ \ 45\ \ \ 55 & 84\ \ \ 77\ \ \ 99 & 22\ \ \ 17\ \ \ 30\\   
10.1889 & 0.2573 & 21.19 & 1.13 & $-$18.48 & 65\ \ \ 42\ \ \ 70 & 13\ \ \ 10\ \ \ 19 & 30\ \ \ 27\ \ \ 36 & 103\ \ \ 93\ \ 131 & 24\ \ \ 14\ \ \ 31 \\ 
10.1899 & 0.0642 & 18.96 & 0.77 & $-$17.07 & $-$\hspace*{0.6cm} &  8\ \ \ \ $\,$7\ \ \ \ $\,$9 & 50\ \ \ 48\ \ \ 53  & 20\ \ \ 19\ \ \ 21 & 8\ \ \ \ $\,$7\ \ \ \ $\,$9\\
14.0025 & 0.2358 & 21.16 & 0.32 & $-$19.11 & $-$\hspace*{0.6cm} & 7\ \ \ \ $\,$6\ \ \ \ $\,$9 & 32\ \ \ 31\ \ \ 35 & 76\ \ \ 68\ \ \ 86 & 18\ \ \ 14\ \ \ 22  \\ 
14.0070 & 0.0495 & 20.73 & 0.24 & $-$16.87 & $-$\hspace*{0.6cm} & 5\ \ \ \ $\,$4\ \ \ \ $\,$6 & 21\ \ \ 19\ \ \ 27 & 39\ \ \ 35\ \ \ 45 & 13\ \ \ 11\ \ \ 18 \\ 
14.0208 & 0.2003 & 21.74 & 0.39 & $-$18.09 & $-$\hspace*{0.6cm}  & 3\ \ \ \ $\,$1\ \ \ \ $\,$6 & 47\ \ \ 40\ \ \ 66 & 72\ \ \ 58\ \ \ 90 & 21\ \ \ 10\ \ \ 27 \\ 
14.0310 & 0.2384 & 20.96 & 0.34 & $-$19.31 & 35\ \ \ 32\ \ \ 37 & 12\ \ \ 10\ \ \ 14 & 33\ \ \ 30\ \ \ 37 & 99\ \ \ 83\ \ 116 & 20\ \ \ 19\ \ \ 26\\ 
14.0377 & 0.2596 & 20.95 & 0.36 & $-$19.50 & 73\ \ \ 65\ \ \ 82 & 21\ \ \ 18\ \ \ 23 & 70\ \ \ 64\ \ \ 76 & 132\ $\,$112\ $\,$156 & $-$\hspace{0.6cm} \\ 
14.0528 & 0.0640 & 20.61 & 0.20 & $-$17.05 & $-$\hspace*{0.6cm} & 11\ \ \ $\,$9\ \ \ 12 & 50\ \ \ 47\ \ \ 55  & 71\ \ \ 65\ \ \ 78 & 14\ \ \ 11\ \ \ 17 \\ 
14.0588 & 0.0992 & 20.58 & 0.34 & $-$17.79 & $-$\hspace*{0.6cm} &  6\ \ \ \ $\,$5\ \ \ \ $\,$7 & 26\ \ \ 24\ \ \ 28  & 49\ \ \ 46\ \ \ 54 & 14\ \ \ 11\ \ \ 17\\ 
14.0621 & 0.2871 & 21.18 & 0.32 & $-$19.55 & 30\ \ \ 27\ \ \ 36 & 3\ \ \ \ $\,$2\ \ \ \ $\,$4 & 15\ \ \ 14\ \ \ 18 & 48\ \ \ 42\ \ \ 56 & 20\ \ \ 17\ \ \ 23 \\ 
14.1039 & 0.0793 & 19.29 & 0.31 & $-$18.64 & $-$\hspace*{0.6cm} & 10\ \ \ \ $\,$9\ \ \ 12 & 44\ \ \ 41\ \ \ 48 & 89\ \ \ 85\ \ \ 92 & 23\ \ \ 21\ \ \ 25\\ 
14.1103 & 0.2092 & 22.33 & 0.31 & $-$17.69 & $-$\hspace*{0.6cm} & 283\ $\,$225\ $\,$360 & 2059 1801 2548 & 1697 1354 2317 & 15\ \ \ \ $\,$7\ \ \ 32  \\
14.1209 & 0.2345 & 20.81 & 0.39 & $-$19.37 & $-$\hspace*{0.6cm} & 1.5\ \ \ \ $\,$1\ \ \ \ $\,$2 & 22\ \ \ 19\ \ \ 24 & 84\ \ \ 76\ \ \ 92 & 14\ \ \ 12\ \ \ 18\\ 
14.1273 & 0.2567 & 22.02 & 0.21 & $-$18.55 & 12\ \ \ 10\ \ \ 16 & 20\ \ \ 15\ \ \ 28 & 60\ \ \ 53\ \ \ 68 & 90\ \ \ 79\ \ 111 & 20\ \ \ 13\ \ \ 26 \\ 
14.1376 & 0.2881 & 21.51 & 0.30 & $-$19.25 & 100\ \ \ 88\ \ 115 & 11\ \ \ \ $\,$9\ \ \ 14 & 17\ \ \ 15\ \ \ 21 & 173\ $\,$139\ $\,$220 & $-$\hspace*{0.6cm}\\ 
14.1425 & 0.2090 & 22.31 & 0.28 & $-$19.11 & 61\ \ \ 55\ \ \ 75 & 10\ \ \ \ $\,$7\ \ \ 13 & 37\ \ \ 30\ \ \ 42 & 102\ \ \ 87\ \ 119 &  9\ \ \ \ $\,$7\ \ \ 15  \\ 
22.0474 & 0.2812 & 21.74 & 0.64 & $-$18.66 & 57\ \ \ 50\ \ \ 65 & 43\ \ \ 35\ \ \ 55 & 170\ $\,$146\ $\,$197 & 355\ $\,$298\ $\,$412 & 53\ \ \ 36\ \ \ 64 \\ 
22.0676 & 0.1409 & 20.58 & 0.51 & $-$18.31 & $-$\hspace*{0.6cm} & 16\ \ \ 13\ \ \ 18 & 55\ \ \ 50\ \ \ 59 & 65\ \ \ 61\ \ \ 74 & 25\ \ \ 21\ \ \ 27 \\ 
22.1013 & 0.2306 & 20.16 & 0.65 & $-$19.71 & 21\ \ \ 17\ \ \ 28 & 9\ \ \ \ $\,$8\ \ \ 10 & 23\ \ \ 22\ \ \ 25 & 47\ \ \ 44\ \ \ 54 & 18\ \ \ 15\ \ \ 23 \\ 
22.1082 & 0.2948 & 21.62 & 0.21 & $-$19.28 & 96\ \ \ 81\ \ 115 & 19\ \ \ 16\ \ \ 23 & 106\ \ \ 97\ \ 118 & 67\ \ \ 60\ \ \ 92 & $-$\hspace*{0.6cm}\\ 
& & & & & & & & & \\ 
\hline 
\end{tabular}
\normalsize 
\caption{\label{next} Spectra analysed in class (A). 
Meaning of the columns (1) CFRS identification
($00.\star\star\star\star$ are galaxies in the 00h field,
$10.\star\star\star\star$, those in the 10h field, etc., RA and DEC are
published in CFRS-II, CFRS-III and CFRS-IV), (2) redshift (see
CFRS-II, CFRS-III, CFRS-IV), (3) apparent magnitude ($I_{AB}=I+0.48$)
(see CFRS-I), (4) photometric colour (see CFRS-I), (5) absolute
magnitude ($B_{AB}=B-0.17$) (see CFRS-VI), (6)-(10) reference-frame
equivalent widths (the second and third values are respectively the
lower and upper limit of the measure) respectively for [O
II]$\lambda3727$, H$\beta$, [O III]$\lambda5007$, H$\alpha$ and [S
II]$\lambda6725$. }
\end{table*} 
\normalsize

\subsection{Measurement of emission-line intensities} 

We have measured the integrated fluxes with the package SPLOT under
IRAF/CL and the MEASURE tool, a code developed in Meudon by D. Pelat
and M. Caillat, an extension of which is presented in Rola \& Pelat
(1994).  The standard deviation of fluxes ($1\ \sigma$) have been
computed with MEASURE; this involves the flux in each pixel of an
emission line and the continuum below the line.  The continuum has
been determined interactively since this allows any defects to be
ignored around the line by considering the general shape over a larger
wavelength range.  The placement of the continuum can be also done
automatically with MEASURE.  If the continuum has a good signal to
noise ($S/N$), the placement by hand is in excellent agreement with
the non-interactive one; we obtain the same result for the fluxes.
The fluxes measured with SPLOT and MEASURE tools are the same within
the flux error range $\pm 1\ \sigma$, and thus secures our
emission-line intensities.  Fluxes have been measured for H$\alpha$,
H$\beta$, [O III]$\lambda5007$, and also for [O II]$\lambda3727$ and
[S II]$\lambda6725$ when these latter lie in the CFRS spectral range
(hereafter we write respectively [O III], [O II], [S II] unless
otherwise is specified). Observed fluxes are given in
Table~\ref{tableflux} and observed EWs in Table~\ref{next}.  In the
case of spectra in class (AB), a flux equal to $1\ \sigma$ above the
continuum has been given as an upper limit to H$\beta$ (and a flux
with no error is written in Table~\ref{tableflux}).

Our spectral resolution separates the [O III]$\lambda4959$ and [O
III]$\lambda5007$ lines, but not the [N II]$\lambda6548$ and
[N II]$\lambda6583$ lines from H$\alpha$.  For a few spectra, [N
II]$\lambda6583$ (whose intensity is roughly three times higher than
[N II]$\lambda6548$) is seen because the line intensity peak is above
the half ma\-xi\-mum of H$\alpha$, and the redshift measured at the
blended H$\alpha$ is higher than the one measured at [S II], or at [O
III].  In such cases, lines have been deblended, and fluxes measured
with the DEBLEND utility under SPLOT using the exact reference-frame
wavelengths of the lines.
 
\subsection{Reddening correction}
\begin{figure}
\caption{\label{histo} Histogram of $A_V$ for spectra 
of class (A). The hatched columns are spectra in sub-class (AB).} 
\end{figure}

Reddening is produced by interstellar extinction along the line of
sight of observed sources.  Interstellar extinction due to our Galaxy
is negligible because the five CFRS fields are located at high
galactic latitude ($\mid b_{II}\ge 45^{o}\mid$), hence it is
mainly intrinsic to the observed galaxy. 

As H$\alpha$ lies always in the CFRS wavelength coverage for galaxies
at $z\leq0.3$, we can calculate the $H\alpha/H\beta$ flux ratio and
correct emission lines for reddening.  The reddening correction $c$ is
given by
$$ \frac{f(H\alpha)}{f(H\beta)} =
\frac{I_{0}(H\alpha)}{I_{0}(H\beta)}\ 10^{-c[I(H\alpha)-
I(H\beta)]},$$ where $I_{0}(H\alpha)/I_{0}(H\beta)=2.86$ for case B
recombination, a density of $100$ cm$^{-3}$, and a temperature of $10\
000$ K (Osterbrock 1989). These are the normal values for H II
regions.  The ratio is negligibly sensitive to temperature.  When $c$
had a negative value, we then considered $c=0$.

We have corrected our fluxes $f(\lambda)$ in the following way:  
$$ \frac{I_{0}(\lambda)}{I_{0}(H\beta)} = 
\frac{f(\lambda)}{f(H\beta)}\ 10^{c[I(\lambda)-I(H\beta)]},  
$$  
where the values of $[I(\lambda)-I(H\beta)]$ have been computed from
Seaton's galactic extinction law (see table 3 in Seaton 1979).  For
$\lambda=6725, 6563, 5007, 4861$ and $3727$, this gives respectively
$-0.3440, -0.3227, -0.0341, 0$ and $0.2553$.  The dif\-fe\-rent extinction
laws behave roughly similarly in the visible range, even though they
diverge in the ultraviolet (see Osterbrock 1989).  Therefore, our
calculations are independent of the choice of an extinction law.
Errors at 1 $\sigma$ level of the reddening corrected line ratios  
are computed as follows: 
$$ \sigma\left(\frac{I_{0}(\lambda_1)}{I_{0}(\lambda_2)}\right)
\simeq \lbrack \frac{\sigma(f(\lambda_1))}{f(\lambda_1)} +
\frac{\sigma(f(\lambda_2))}{f(\lambda_2)} \rbrack \times
\frac{f(\lambda_1)}{f(\lambda_2)},$$ where the 1 $\sigma$ flux errors
have been measured as described in Section 3.1.
The reddening parameter $A_{V}$ ($V$ for visual) is:
$$ A_{V} = \frac{c R}{1.47}\  mag, R=3.2 .$$  

In Figure~\ref{spectres}, we note that stellar absorption is seen at
H$\alpha$ and H$\beta$ for spectra (AB).  
The H$\beta$ emission cannot be resolved with our low resolution. The
effect for H$\alpha$ emission is less noticeable because of the line
strength.  In these cases, H$\alpha$ and H$\beta$ fluxes are
underestimated. Therefore, the observed $H\alpha/H\beta$ ratio is an
upper limit since H$\beta$ stellar absorption is stronger than the
H$\alpha$ one.
$H\alpha/(3\times H\beta)$ would be a better estimate, and $A_V$ would
be then $0.5$ magnitude smaller.  As we mentioned above, if the
intensity of [N II] is strong, its contribution to H$\alpha$ was
measured, otherwise H$\alpha$ includes [N II] for part of spectra (A),
and this makes the observed $H\alpha/H\beta$ ratio larger.  We note
that the stellar absorption and the presence of the blend [N II] at
H$\alpha$ have opposite effects, and thus we cannot determine whether
the true value of H$\alpha$ is larger or smaller.

In the histogram of $A_V$ (see Figure~\ref{histo}), hatched columns
distinguish spectra (AB) for which $A_V$ is likely overestimated.
Reddening parameters are listed in Table~\ref{tableflux}.  Four 
spectra (AB) have $A_V > 4$ mag, however we notice that spectra (AB) do not
have systematically large $A_V$.  Our reddening range, $0 < A_V < 2-3$
mag, is consistent with those of nearby H II-region samples (Oey \& 
Kennicutt 1993, Kennicutt {\it et al.} 1989). The average reddening
for $A_V < 4$ mag is $1.52$ mag; higher than that from Oey \&  
Kennicutt (1993), $ \langle A_V = 1$ mag $\rangle $.  
$A_V$ may be overestimated because of stellar absorption mostly 
underlying H$\beta$, and [N II] weak lines blended with H$\alpha$. 
We investigate in more detail stellar absorption effects in Section 6.

\section{Emission-line diagnostic diagrams}  
\begin{figure*}
\caption{\label{loca} Diagnostic diagrams with a compilation of 
emission-line 
galaxy samples taken from the literature (see Table 4).  
Lines are empirical separations based
on this compilation, between H II galaxies (left), Seyfert 2 galaxies (top
right) and LINER galaxies (bottom right).}
\end{figure*} 
\begin{table*}
\begin{center}
\begin{tabular}{||c|c|c||}
\hline 
Objects & Diagrams & References \\
\hline  
\hline 
Seyfert 2  & 1 \& 2 & Shuder \& Osterbrock, 1981, ApJ, {\bf 250}, 55 \\
 Seyfert 2 &  1 \& 2 & Koski, 1978, ApJ, {\bf 223}, 56 \\
 Seyfert 2 & 1 \& 2  & Phillips, Charles \& Baldwin, 1983, ApJ, {\bf 266}, 485 \\  
 LINERs & 1 & Keel, 1983, ApJ, {\bf 269}, 466 \\
 LINERs & 1 \& 2 & Ferland \& Netzer, 1983, ApJ, {\bf 264}, 105 \\ 
 Seyfert 2 & 1  &  Veilleux \& Osterbrock, 1987, ApJS, {\bf 63}, 295 \\
 Narrow Emission Line Galaxies (NELG)      &  1 &   \\
 Starburst galaxies (SBG) & 1  &    \\
 H II Regions & 1 \& 2 & Pena, Ruiz \& Maza, 1991, A\&A, {\bf 251}, 417 \\
 H II Regions & 1  &  McCaugh, 1991, ApJ, {\bf 380}, 140 \\
 Blue Compact Galaxies (BCG)  & 1 \& 2 & Izotov, Thuan \& Lipovetsky, 1994, ApJ, {\bf 435}, 647 \\
\hline 
\end{tabular}
\end{center}
\caption{\label{references} Table of the local galaxy sample. 
   We use the published data either for both emission-line diagrams,
  or for a single one depending on the available emission lines.
  The diagram 1 refers to the [O III]/H$\beta$ {\it vs.}
  [S II]/H$\alpha$ diagram, and the diagram 2 refers to the
  [O III]/H$\beta$ {\it vs.} [O II]/H$\beta$
  diagram.}
\end{table*} 

Emission-line diagnostic diagrams provide a reliable classification of
narrow emission-line spectra according to the main ionization source
responsible for these lines. With the [O III], [S II] and [O II] lines, we
can use the [O III]/H$\beta$ {\it versus} [S II]/H$\alpha$ and [O
III]/H$\beta$ {\it versus} [O II]/H$\beta$ diagnostic diagrams.

The [O III]/H$\beta$ {\it vs.} [S II]/H$\alpha$ diagnostic diagram (see VO)
offers the significant advantage of involving ratios of lines with
small wavelength separation, minimizing uncertainties introduced in
flux calibration and reddening correction.  As shown in
Figure~\ref{loca}, it clearly separates H II galaxies from active
galaxies.  Active galaxies can be separated into Seyfert 2 and LINERs,
the latter having a smaller excitation degree ($\log [O\ III]/H\beta
\leq 0.5$) than the former.  The separations drawn between H II, 
Seyfert 2 and LINER galaxies are empirical, based on the
distribution of a compilation of emission-line galaxy samples taken
from the li\-te\-ra\-ture (the references for these are given in
Table~\ref{references}).  In the [O III]/H$\beta$ {\it vs.} [O II]/H$\beta$
diagnostic diagram, the different sources are also well separated in
contrast with the [O III]/H$\beta$ {\it vs.} [O II]/[O III] diagram used in
earlier works (see Baldwin, Phillips \& Terlevich 1981, hereafter
BPT). The [O II]/H$\beta$ ratio depends on reddening.  The single
oxygen element involved in this diagnostic is an advantage: the
distribution of data and photoionization models are less dependent on
metallicity of other elements (see Section 5).

\subsection{The [O III]/H$\beta$ vs. [S II]/H$\alpha$ diagram}
\begin{figure*}
\caption{\label{dia} Diagnostic diagrams with the emission-line galaxies 
of class (A) (see Section 2). The error bars on the points show 
$1\ \sigma$ errors.  Line ratios are dereddened. The diagram on the
left panel does not depend on reddening conversely to the one on the
right panel (see Figure 5). Empirical limits between H II, 
Seyfert 2 and LINER galaxies are taken from Figure 3.}
\end{figure*}

In Figure~\ref{dia} (left panel) we present our reddening corrected
data with the respective $1\ \sigma$ error bars in a logarithmic
scale.  We emphasize that the ratios involve lines having
approximately the same reddening corrections, hence values plotted in
this diagram depend very little on this correction.  The same
empirical boundaries as shown in Figure~\ref{loca} for the locus of H
II, Seyfert 2 and LINER galaxies are drawn.  Solid points represent
spectra for which [O II] and [S II] lie within the observed wavelength
range.  Open points represent spectra for which [S II] lies within the
observed wavelength range, but [O II] does not.  The two types of
points are distributed similarly in the diagram.  Solid points are
data which can be used in our second diagnostic diagram, and thus
allow a comparison between both diagrams (see Section 5).  In
addition, different symbols have been chosen as follows: circles are
spectra for which all line intensities could be measured (mainly class
(A) without the sub-class (AB)), triangles and squares represent
spectra for which $1\ \sigma$ limit has been measured respectively for
the H$\beta$ intensity only (sub-class (AB)) and for the [S II]
intensity only.  In a few cases the $(S/N)$ of [S II] is too low: the
line intensity is effectively faint and the $(S/N)$ of the continuum
around the line is not good.

The sub-class (AB) (triangles) is not distributed dif\-fe\-ren\-tly in the
diagram compared to circles, although pre\-fe\-ren\-tial\-ly
located on the right (high [S II]/H$\alpha$).  Squares do not present
a particularly different distribution, just have a low [S
II]/H$\alpha$.  We have chosen different symbols to indicate that some
spectra exhibit emission lines with a $S/N<2$, and some spectra have
excellent quality.  We conclude that there is no particular location
in the diagram for good data (small error bars) and the rest.

The unexpected result is the significant fraction of data lying in the
active galaxy area, especially in the Seyfert 2 one.  We notice also
that these data do not present a high excitation degree ($[O\
III]/H\beta \gg 0.6$) such as the traditional Seyfert 2 galaxies.
They are in the low excitation part of the locus of Seyfert 2 galaxies
($[O\ III]/H\beta \simeq 0.5-06.$), but not as low as LINER galaxies
($[O\ III]/H\beta < 0.4$).  Globally, we find that the domain occupied
by the CFRS emission-line galaxies is very restricted with respect to
the one occupied by all objects in Figure~\ref{loca}. Our
emission-line galaxies are distributed between the higher ionization
bound of H II galaxies and the lower ionization area of Seyfert 2
galaxies.

Underlying absorption at Balmer lines would make the observed $[O\
III]/H\beta$ and $[S\ II]/H\alpha$ ratios larger, and true values
should be shifted towards the bottom-left of the dia\-gram, whatever the
reddening since these ratios are not dependent on it. We note that
H$\alpha$ emission is less con\-ta\-mi\-na\-ted by stellar absorption than
H$\beta$ emission, but can be contaminated by weak [N II] lines (see
Section 3.1).  However stellar absorption would shift data more
towards the bottom than the left of the diagram into the LINER area,
and thus still lying in the active galaxy area.  Besides, changing
$\log ([S\ II]/H\alpha)=-0.6$ to $-0.4$ would imply a H$\alpha$
intensity $1.6$ times lower, or a [S II] intensity $1.6$ times higher.
This represents a large shift in flux of at least $10\ \sigma$ from
most of our flux measures.

\subsection{The [O III]/H$\beta$ vs. [O II]/H$\beta$ diagram} 

In this diagram (Figure~\ref{dia}, right panel), the same symbols have
been chosen as in Section 4.1, but now open symbols are spectra for
which [O II] lies within the observed wavelength range, and [S II]
does not.  Line ratios are corrected for reddening.  As previously,
data with different symbols do not show a particular distribution in
the diagram, the sub-class (AB) still being preferentially on the
right (high [O II]/H$\beta$).  We find that a significant
fraction of data has $\log\ ([O\ II]/H\beta)$ and $\log\ ([O\
III]/H\beta)$ greater than $0.6$; this is too high for H II
galaxies. As in the previous diagram, we notice that the distribution
of the CFRS emission-line ga\-la\-xies is restricted compared to the
objects plotted in Figure~\ref{loca}.
 
This diagram depends on reddening because of the [O~II]/H$\beta$
ratio.  Non-dereddened line ratios are plotted in
Figure~\ref{nondered}.  Reddening correction moves data towards larger
[O~II]/H$\beta$ (Figure~\ref{dia}).  We notice that for the
uncorrected data, a significant fraction still lies in the active
galaxy area.  However we emphazise that line ratios must be dereddened
for a consistent determination of the ionization sources.  Underlying
absorption at H$\beta$ would make the observed [O~II]/H$\beta$ and [O
III]/H$\beta$ ratios larger, and true values should be shifted to the
bottom-left in Figure~\ref{dia}.  This leads to a non-straightforward
situation, because correcting for reddening (data move towards the
right) has the opposite effect as correcting for stellar absorption
(data move towards the bottom-left), and the dominant effect depends
on each object.  
Therefore, because of possible stellar absorption, the
dereddened diagnostic diagram is used as a support to the previous
one.  However if the stellar absorption can be measured, this diagram
is very useful to distinguish emission-line galaxies of different
ionization sources as shown in Figure~\ref{loca}.

We conclude that a significant fraction of galaxies is also seen in
the active galaxy area in this second diagnostic diagram. Indeed, few
non-reddened data should have sufficient strong stellar absorption to
move into the H II galaxy area. For example, changing $\log\ [O
II]/H\beta=1$ to $0.5$ would imply a H$\beta$ intensity $\sim 7$ times
higher. As we will show in Section~6, some spectra in the active
galaxy part do not seem to exhibit a stellar absorption at H$\beta$ as
strong as that.
\begin{figure}   
\caption{\label{nondered} [O III]$\lambda5007$/H$\beta$ {\it vs.}  
[O II]$\lambda3727$/H$\beta$ diagnostic dia\-gram with 
non-dereddened lines. The error bars on the points show
$1\ \sigma$ errors. Comparison with Figure 4 indicates the direction 
in which points move as reddening corrections are applied.}  
\end{figure}

\section{Determination of H II galaxy boundaries} 

The location of H II and active galaxies in diagnostic diagrams is
generally determined by the distribution of observational data and the
loci of some photoionization models which try to reproduce the line
ratios observed (see BPT, VO). However, until now, no work based on
theoretical mo\-dels has established the maximal extent of the location
of sources photoionized by OB stars in these diagrams.  For our
analysis, the determination of these limits is crucial to understand
whether all our emission-line ratios can be reproduced by OB star
ionization sources.

The main difference between H II and active galaxies comes from the
ionizing spectrum.  Indeed, the gas ionization is produced by
ultraviolet photons emitted by OB stars in H~II galaxies, and by a
significant fraction of X-ray photons in active galaxies.  The
H$^{0}$, He$^{0}$, He$^{+}$ and other ion cross sections decrease
rapidly with increasing photon energy, hence X-ray photons tend to
penetrate deeper than ultraviolet photons in the nebula.  Thus a large
partially-ionized region is produced where very energetic free
electrons coexist with neutral atoms and less ionized ions such as
O$^{0}$, O$^{+}$, S$^{+}$, N$^{0}$, and N$^{+}$.  Therefore this 
 region is more extended in active galaxies than in H
II galaxies where the transition between ionized gas and neutral gas
is steep.  The intensities of the lines emitted by these atoms and
ions will then reflect this difference; this is the main idea on which
emission-line dia\-gnos\-tic diagrams are based. In this section, we
present our photoionization models, and their behaviour in the
diagnostic diagrams we used. Finally we compare our data to the upper
models found for OB star ionization sources.

\subsection{Photoionization models}
\begin{figure*}
\caption{\label{claudia1} Behaviour of the photoionization models in 
both diagnostic diagrams as a function of the source effective
temperature, $T_{\rm eff}$, for a fixed metallicity $Z= Z_{\odot}$. 
We can observe the variation of each intensity line ratio with
$T_{\rm eff}$ ranging from $40\; 000$ K to $60\; 000$ K.}
\end{figure*}
\begin{figure*}
\caption{\label{claudia2}Behaviour of the photoionization models in both 
diagnostic diagrams as a function of the metallicity, $Z$, for a fixed
source effective temperature, $T_{\rm eff} =$ $60\; 000$ K.  The
metallicity ranges from $0.1 \times Z_{\odot}$ to $2 \times Z_{\odot}$
We notice that in both diagrams, the upper curve defines the envelope
for the maximum extent of the locus of H II galaxies.}
\end{figure*}

We have calculated a grid of steady-state spherically symmetric H
II-region models, using the photoionization code PHOTO (Stasi\'nska
1990).  The grid allows us to obtain in several diagnostic diagrams
the upper limit of sources photoionized by OB stars (see Rola 1995).
We have constructed the grid by carefully considering wide ranges for
the physical parameters typical of H II regions; this ensures finding
the whole extent of these regions in the diagrams.

A H II galaxy presents a high degree of complexity re\-la\-ti\-ve to a
single H II region.  They differ in their dimensions, in the gas and
stellar content.  Moreover, H II galaxies are a mixture of gas
excitation mechanisms, whereas gas photoionization is the dominant
process in H II regions.  Since emission lines observed in H II
galaxies are mainly produced by the H II regions contained within,
using H II-region mo\-dels in the analysis of H-II galaxy emission-line
ratios is reasonably accurate.  We describe in the following
paragraphs the construction of our grid of models.  

The effects of dust on the ionizing spectra are still controversial
(see Mathis 1986, Shields \& Kennicutt 1996), and seem to depend on
unknown grain optical properties. Thus we simply choose dust-free
models in which the existence of dust is considered only at the level
of depletion (che\-mi\-cal elements in the solid state); this is actually
the main dust effect expected on emission-line ratios.  Models are
ionization bounded, {\it i.e.} the limits of photoionized regions are
defined by the strength of ionization sources, and not by the gas
mass. This choice is consistent with most H II-region observations
(see {\it e. g.}, McCall {\it et al.} 1985).  To reproduce the content
of H II-region ionizing sources, a star cluster is often taken with a
certain distribution of stellar masses and effective temperatures
({\it e.g.} McGaugh 1991).  Nevertheless, if the hottest stars are
assumed to dominate the total flux of io\-ni\-zing photons,
a good approximation is to consider a single type of ionizing
spectrum. The latter depends only on stellar effective temperature and
metallicity.  For the distribution of ionizing spectra we use the
$\log \ g = 5$ model at\-mos\-phe\-res of Kurucz (1992) taking abundances
consistent with the ones in the modelled nebula.
We have also constructed a grid of models considering a black body
distribution for ioni\-zing spectra.  Hereafter the described
photoionization mo\-dels are calculated with Kurucz (1992) model
atmospheres unless otherwise mentioned.

There are five main parameters driving the photo\-io\-ni\-zation models: the
hydrogen density, $n_{\rm H}$, the metallicity, $Z$, the number of
ionizing stars, $N$, the filling factor,$f$, and the source
effective temperature, $T_{\rm eff}$. 

We have tested several values of gas density within the range of
typical H II-region values.  We find that the variation of the gas
density has a negligible effect on the emission-line intensity ratios.
This is because the gas density of H II regions is generally less 
than the critical density for collisional de-excitation for most
emission lines.  Thus in this analysis, we restrict our choice to a
constant density, $n_{\rm H}= 10$ cm$^{-3}$. 
 
We consider four metallicities: $2\ Z_{\odot}$, $Z_{\odot}$, $0.25\
Z_{\odot}$ and $0.1\;Z_{\odot}$, each metallicity being defined by a
set of element abundances $X/H$.  We determine $X/H$ from the
$O/H$-dependent expressions appropriate to H II regions (McGaugh 
1991).  We take for $O/H$ the solar oxygen abundance value given by
Anders and Grevesse (1989)\footnote[2]{$(O/H)_{\odot} = 8.51
\times 10^{-4}$}. The set of ``solar abundances'' $Z_{\odot}$ actually
contains only one solar abundance value, $(O/H)_{\odot}$. Thus,
 $x\ Z_{\odot}$ means that the $(O/H)$ abundance is taken as
$x\times (O/H)_{\odot}$ in the $(X/H)$ expressions.  We consider 
also that magnesium, silicium and iron are depleted
 by $10\%$ (Stasi\'nska 1990).

The ionization parameter $U$ defines the degree of io\-ni\-za\-tion of the
gas.  Globally, it represents the average ratio between the flux of 
hydrogen ionizing photons and the gas density, and is given by
\begin{equation}
U= Q_{\rm H}/( 4 \pi R^{2} n_{H} c), 
\label{eq-grille:1}
\end{equation}
where $Q_{\rm H}$ is the number of hydrogen ionizing photons, $R$ is
the Str\"omgren radius and $c$ is the light speed.  The ionization
parameter can be also related to the filling factor and to the number
of ionizing stars 
\begin{equation} U \propto \left( n_H f^2
N Q^{\star}_H \right) ^{\frac{1}{3}},
\label{eq-grille:2}
\end{equation}
where $Q^{\star}_H $ is $Q_{\rm H}$ for a single star. We vary $U$
within the range $0.2 - 10^{-6}$, approximately.
 
We vary $T_{\rm eff}$ within the range $3 - 6 \times \ 10^{4}$ K.  The
$6 \times 10^{4}$~K model atmospheres are generated from the $5 \times
10^{4}$~K Kurucz model atmospheres; we derive a scaling factor from the
ratio of the $5 \times 10^{4}$ K and $6 \times 10^{4}$ K blackbody
mo\-dels, and multiply the $5 \times 10^{4}$ K Kurucz model fluxes by
this factor (Leitherer, private communication).  According to the
Maeder (1990) stellar models, $T_{\rm eff} = 6 \times \ 10^{4}$ K
cor\-res\-ponds to a 120 M$_{\odot}$ main sequence star with $Z= 0.1\
Z_{\odot} $. Furthermore, from the mass-temperature-metallicity
correlation obtained from Maeder models (see figure 2 in McGaugh
1991), the true maximum effective temperature of a H II-region
io\-ni\-zing star is somewhere between $5 \times \; 10^{4}$ K and $6
\times 10^{4}$ K.  After this work was completed, Stasi\'nska \&
Leitherer (1996) determined that in a giant HII region ionized by a
cluster of stars with masses distributed with a Salpeter initial mass
function and an upper cutoff at $100$ M$_{\odot}$, the equivalent
effective temperature is between $4.5 \times 10^{4}$~K and $5 \times
10^{4}$~K depending on the metallicity.  A cutoff at $120$ M$_{\odot}$
produces negligible differences.  The maximum effective temperature
for O stars may be lower than $60\ 000$ K, hence this value must be
considered as an upper limit for our models.
\begin{figure*}
\caption{\label{modloca} The maximum boundary 
for gas photoionization by OB stars in both diagrams is defined by the
solid curve (corresponding to the models with Kurucz (1992) model
atmospheres with $T_{\rm eff} =$ $60\ 000$ K and $Z= 0.25 \
Z_{\odot}$).  The dotted curve is plotted for comparison and
corresponds to the same models but for $T_{\rm eff} =60\ 000$ K and
$Z= 0.1\ Z_{\odot}$. To study the effect of changing the ionizing
spectra we present the models calculated with a blackbody spectra for
$T_{\rm eff} =60\ 000$ K and $Z= 0.25\ Z_{\odot}$ (dashed curve).  The
dot-dashed curve is the observational H II-region sequence from
McCall {\it et al.} (1985).}
\end{figure*}
\begin{figure*}
\caption{\label{compa} Comparison of the data with the limits from the 
models in the two diagnostic diagrams. The solid curves are the
envelope of H~II galaxies for $T_{eff}=60\ 000$ K (same as in Figure
8), and the dotted curves are the same envelope for $T_{eff}=50\ 000$
K (see Section 5.2). Circles represent spectra with both [O~II] and [S~
II] lines in the observed spectral range. Squares represent spectra
with either \mbox{[S~II]}, or [O~II] within the spectral range. Solid points
are data in the active galaxy area, open points are data in the H II
galaxy area, open points with a cental dot are data between areas. This
classification has been done in the diagram on the left with the
uppest envelope, and carried directly to the diagram on the right.}
\end{figure*}

\subsection{Behaviour of the models} 
 
Varying the ionization parameter, $U$, the stellar effective
temperature, $T_{\rm eff}$, and the metallicity, $Z$, allows study of
the global behaviour of the models in the diagnostic diagrams.
Figure~\ref{claudia1} shows the behaviour of the models varying with
$T_{\rm eff}$ for $Z = Z_{\odot}$ in the two diagnostic diagrams used
in this paper.  The solid lines connect the models with the same
number of ionizing stars, $N$ and filling factor, $f$.  $U$ slightly
diminishes along the line towards the bottom of the diagram.  The
dashed curves connect the models with the same $T_{\rm eff}$, and with
$U$ increasing along the curves towards the left because of the
variation of $N$ and $f$.  We see that an increasing $T_{\rm eff}$
produces harder stellar photons, hence an increasing proportion of
O$^{++}$ ions; this gives higher [O~III]/H$\beta$ ratios.
Figure~\ref{claudia2} shows the behaviour of the models with $Z$ for
$T_{\rm eff} = 60 \ 000$ K.  The solid lines correspond to the
variation of line intensity ratios with $Z$ (where $N$ and $f$
are constant, and $U$ slightly varies). The dashed curves correspond
to the variation of $U$ increasing towards the left.  An increase in
$Z$ (downwards the solid line) represents an increase in the number of
coolants in the gas; this causes a diminution in the electronic
temperature $T_{\rm e}$ and a significant decrease in the intensity of
the forbidden lines.

In each diagnostic diagram, the envelope of the ma\-xi\-mum extent of
sources photoionized by OB stars is defined by the upper limits of our
models.  In Figure~\ref{modloca}, the envelope (solid line) is plotted
for each diagram with the compilation of observational data taken
from the literature (see Table~\ref{references}).  In the models
defining the envelope, $U$ varies from $\sim7.4 \times \ 10^{-2}$ to
$\sim 2.1 \times \ 10^{-6}$ respectively from the top left to the
bottom right of the diagrams.  In the [O III]/H$\beta$ {\it vs.} [O
II]/H$\beta$ diagram, the envelope corresponds to models with
$T_{\rm eff} = 60\ 000$ K and $Z = 0.25\ Z_{\odot}$.  In the [O
III]/H$\beta$ {\it vs.} [S II]/H$\alpha$ diagram, the envelope corresponds
to the models with $T_{\rm eff} = 60 \ 000$ K and $Z = 0.25 \
Z_{\odot}$ if $\log \ ([O\ III]/H\beta)> 0$, approximately and $Z =
Z_{\odot}$ if $\log ([O\ III]/H\beta) < 0$, approximately.  Models
with $T_{\rm eff} = 60\ 000$ K and $Z = 0.1\ Z_{\odot}$ (dotted line)
are also shown.  We note that the corresponding envelope for models
with a blackbody spectrum as ionizing source (long-dashed line)
underestimates the emission-line intensity ratios relative to the
models with Kurucz model atmospheres.  As re\-fe\-ren\-ce, we
have drawn the observational H II-region sequence of McCall {\it et
al.}  (1985) (dot-dashed line).

We emphasize that some uncertainties associated with the predicted
S$^{+}$/S$^{++}$ ratio in H II-region photoionisation models could
affect the predicted [S II]/H$\alpha$ ratio, and hence the location of
the envelopes. The uncertainties are related to the incompleteness of
atomic data (Dinerstein \& Shields 1986, Garnett 1989).
Nevertheless in Figure~\ref{modloca}, the envelope defined in the [O
III]/H$\beta$ {\it vs.} [S II]/H$\alpha$ diagram (the one that should be the
most affected by these uncertainties) agrees remarkably well with the
compilation of observational data, noting that the classification of
these data has been done by other authors. The same remark can be
made for the other diagnostic diagram. Thus this assures that these
uncertainties would not significantly affect our models, and hence the
location of the envelope of the maximum extent of H II galaxies in
these diagrams.

\subsection{Comparison of the data with photoionization models} 

In this section, we compare the distribution of our data in the two
emission-line diagnostic diagrams with the theo\-re\-ti\-cal upper
limits of the location of H II galaxies.  As shown in Section 4, there
are no differences in the distribution bet\-ween the data represented
by various symbols in Figure~\ref{dia}, hence we keep two symbols to
compare easily the diagnostic diagrams as shown in Figure~\ref{compa}.
Circles are spectra with [O II], [O III] and [S II] lines within the
CFRS wavelength range; these data are plotted in both diagnostic
diagrams.  Squares are spectra with [O III] and either [O II], or
[S~II] within the CFRS wavelength range; the data are plotted in the
diagnostic diagram corresponding to the second line observed.  The
classification of our data is based only on the envelope of the H II
galaxy location in the [O III]/H$\beta$ {\it vs.} [S~II]/H$\alpha$
diagram (see Figure~\ref{compa}).  Solid and open points represent
respectively data lying within the active and H II galaxy area.  Open
points with a central dot are data considered as intermediate between
areas.

Comparing circles between the two diagnostic diagrams,
the data classified as active galaxies from the [O
III]/H$\beta$ {\it vs.} [S II]/H$\alpha$ diagram lie also in the active
galaxy area in the [O III]/H$\beta$ {\it vs.} [O II]/H$\beta$ diagram. Hence
the observed [O~II]/H$\beta$ ratios are consistent with the observed
[S~II]/H$\alpha$ ratio; this is what is expected from models and from
observed emission-line galaxies.

We find for 14 galaxies that their line ratios are definitely outside
the upper limit of the location of H II ga\-la\-xies in both diagnostic
diagrams.  This result implies that the main gas ionization source of
these galaxies cannot be only massive O stars, and must involve harder
ionization sources.  Their line ratios are consistent with observed
active ga\-la\-xies such as Seyfert 2 or LINERs.  However from the present
study, we cannot confirm if they are such ga\-la\-xies, or if they are of
a different nature.  We emphazise that ionization by post-AGB stars
from an old stellar population (see Binette {\it et al.} 1994) seen in
early type galaxies is ruled out since it would produce much smaller
emission-line e\-qui\-va\-lent width for the (H$\alpha$ + [N II]) line
complex than observed in our galaxies (all above $20\ \AA$), as well
as red colours instead of blue. Ionization by extreme Wolf-Rayet stars
could reproduce some of our ``active galaxy'' line ratios (Terlevich
{\it et al.} 1993). However, stellar model atmospheres for these stars
are uncertain because of a lack of comparison with observational data
in the ionizing UV, and thus one cannot assert the above possibility. 

\section{Underlying stellar absorption} 

In this section, we discuss possible effects of stellar absorption
which we are unable to measure accurately because of our low
spectral resolution ($40\ \AA$). This is the main effect that could
significantly affect our results. We address this problem by analysing
combined spectra, and through an equivalent width ratio diagram.

\subsection{Combined spectra} 
\begin{table*}
\begin{center}
\small
\begin{tabular}{|c|cccc|} 
\hline 
& & & & \\
Combined & A$_{V}$ & log [O II]3727/H$\beta$ & log [O III]5007/H$\beta$ 
  & log [S II]/H$\alpha$ \\ 
spectrum & (mag) & & & \\
\hline
& & & & \\ 
H II galaxy & 1.114 & 0.673 ($+0.016-0.017$) & 0.523 ($+0.011-0.011$) & 
 $-0.894$ ($+0.031 -0.034$) \\   
Active galaxy & 2.737 & 1.252 ($+0.026-0.028$) & 0.533 ($+0.021-0.023$)  & 
 $-0.459$ ($+0.020-0.021$) \\
& & & & \\
\hline 
\end{tabular}
\normalsize
\end{center}
\caption{\label{tabcomb} Line intensity ratios of the combined spectra 
plotted in Figure 10. Errors are at $1\ \sigma$ level.}     
\end{table*}
\normalsize
\begin{figure}
\caption{\label{combine} Combined spectrum of objects in the H II  
galaxy area (top), and in the active galaxy area (bottom).}
\end{figure}

We produce a combined spectrum from objects in the active galaxy area,
and another from those in the H II galaxy area (see
Figure~\ref{combine}).  We only consider the spectra of class (A)
without its sub-class (AB), and for which both [O II] and [S II] lines
are inside the CFRS spectral range.  The line ratios from the two
combined spectra are listed in Table~\ref{tabcomb}.  The combined
``active galaxy'' spectrum exhibits more intense [O II] and [S II]
line intensities than the combined ``H~II galaxy'' spectrum. The
former exhibits Ca H\&K and H$\delta$ lines in absorption, tracers of
the underlying stellar po\-pu\-la\-tion.  We notice also that H$\gamma$
appears at the same level as the continuum. Thus, if there is any
strong stellar absorption under the Balmer lines for the considered
spectra, it is not sufficiently strong to render H$\gamma$ in
absorption. This suggests that stellar absorption for the considered
spectra must be weak, since H$\gamma$ is stronger than H$\beta$ in
absorption, and is about twice fainter than H$\beta$ in emission.  The
absence of absorption features such as MgI($5183\ \AA$) in both
combined spectra supports this conclusion.  The absence of H$\gamma$
in the ``active galaxy'' spectrum may indicate that the underlying
stellar absorption is higher in this spectrum than in the ``H~II
galaxy'' one.  Hence the true value of H$\beta$ could be slightly
higher for the former ones, implying more objects in the LINER area
than observed.

\subsection{Analysis of the stellar absorption under the Balmer lines} 

Stellar absorption under the Balmer lines is
surely present in some spectra, especially in those of the sub-class
(AB).  
We test the effect of underlying stellar absorption by u\-sing the
EW([O III]/EW(H$\beta$) {\it vs.}  EW([S II])/EW(H$\alpha$)
dia\-gnos\-tic diagram.  The observed EWs are given in
Table~\ref{next}.  This diagram is roughly equivalent to the [O
III]/H$\beta$ {\it vs.} [S~II]/H$\alpha$ diagram, since the line
ratios involve lines with small wavelength separation with
approximatively the same con\-ti\-nuum.  It does not depend on reddening,
as we stressed in Section 4, and thus enables us to test {\it
directly} any underlying stellar absorption effect by adding a given
stellar absorption correction, called EW$_c$, to rest-frame EWs of
H$\alpha$ and H$\beta$. Normally, EW$_c$ should be smaller for
H$\alpha$ than for H$\beta$.  Nevertheless, for simplicity we use the
same value: $0\ \AA$ (equivalent to our previous analysis), $2\ \AA$
(usually used in the literature for H II regions), and $5\ \AA$ (see
Kennicutt 1992).  The results are shown in Figure~\ref{eqw}.  In the
case of $EW_c=0\ \AA$, 14 objects (8 in the sub-class (AB)) remain in
the active galaxy area, as in the respective line intensity ratio
diagram.  In the case of $EW_c=2\ \AA$, 10 objects remain in the
active galaxy area (6 in the sub-class (AB)), and in the case $EW_c=5\
\AA$, only 4 objects (3 in the sub-class (AB)) remain in it.  It is
unlikely, as already discussed, that spectra in class (A), but not
classified (AB), have strong stellar absorptions. A value of $2\ \AA$
should be a good estimate for these objects, thus 4~objects lie in the
active galaxy area. If we consider $5\ \AA$ for spectra (AB), 3
objects still lie in the active area.  In total, 7 objects remain in
the active galaxy area if we consider these most likely extreme
stellar absorption effects.  We note that it would be unreasonable to
apply more extreme EW$_c$ to all spectra (AB).  This minimum number is
already extreme: (a) H$\alpha$ stellar absorption is less strong than
H$\beta$ one, (b) any presence of weak broad line underlying Balmer
lines would imply that the true value of the narrow component of
H$\alpha$ is lower, (c) H$\alpha$ may be blended with weak [N II]
lines, (d) the upper envelope at $T=60\ 000K$ of H II galaxies is an
extreme boundary (see Section 5).  We conclude that at least 7 sources
need harder photoionization sources than massive O stars, and have
line ratios consistent with active galaxies.
\begin{figure} 
\caption{\label{eqw} Effect of the stellar absorption underlying the 
Balmer lines.  The correction $EW_c$ is $0\ \AA$ for the solid points,
$2\ \AA$ for the open points, and $5\ \AA$ for the starred points. The
same sample of galaxies is used for each $EW_c$, and open points are common
to both figures.  Lines are the same as in Figure 9, the solid one
being the upper limit of the H II galaxy locus.}
\end{figure} 

\section{Statistics}

The class (A) consists of 74 emission-line spectra from a low-$z$
sample of 138 galaxies. We were able to analyse 28 spectra in the two
diagnostic diagrams ({\it i. e.}  containing both the [O II] and [S
II] lines in the CFRS spectral range). We have not analysed 7 spectra
for which an emission line is affected by zero order (see CFRS-II),
and 17 affected by sky lines.  We consider an emission line affected
by a sky line when the profile is clearly deformed; this happens when
the emission and sky line are at the same wavelength, or close by
$\sim \pm 50 \AA$.  There is no reason for these spectra to have a
different distribution in the diagrams.  We find that $10$ of the $28$
spectra have line ratios above the upper theoretical limit of the H II
galaxy location, {\it i.e} $(10/28)\times (74/138) \simeq 19 \%$ of 
all galaxies at $z\leq 0.3$.  
Including all the analysed objects (42 in total) in the [O
III]/H$\beta$ {\it vs.} [S II]/H$\alpha$ diagram, we obtain about the
same percentage, $(14/42)\times (74/138) \simeq 19 \%$.
Accounting for the maximum likely effects of stellar absorption, the
percentage is reduced to $(7/42)\times (74/138) \simeq 9\%$.

Analysis of the galaxies for which no redshift was measured, indicated
that $65$ should have the same distribution as the $591$ identified
ones, and that the rest are probably at $z>0.5$ (see CFRS-V).  Thus,
$23.3\% (=138/591)$ of the $65$ spectra should be at $z\le 0.3$, {\it
i.e.} 15 galaxies.  We assume that these galaxies belong to the class
(C), because the presence of an emission line would have helped to 
measure a redshift.  Accounting for this incompleteness, 
the previous percentages become as $17.3\%$,
$(10/28)\times [74/(138+15)]$ and $8\%$, $(7/42)\times [74/(138+15)]$.

The present statistics are lower than the ones presented in Tresse
{\it et al.} (1995), because in this earlier work we did the statistics
with an empirical delimitation between H II and active galaxy areas
from VO, and not with the theoretical boundaries of H II galaxies
as determined in the paper. 
\begin{figure*}
\caption{\label{zI} $I_{AB}$ apparent magnitude {\it versus} the redshift $z$ 
for the $138$ galaxies classified in the spectral classes (A), (B) and
(C) as defined in Section 2 (left), and for galaxies in class (A)
classified in Figure 9 (right).}
\end{figure*}
\begin{figure*}
\caption{\label{color} $(V-I)_{AB}$ colour {\it versus} the redshift $z$ for 
the $138$ galaxies classified in the spectral classes (A), (B) and (C)
as defined in Section 2 (left), and for galaxies in class (A)
classified from Figure 9 (right). The curves represent the spectral
colours for elliptical (E), spiral (Sb) and late spiral or irregular
(Sd) galaxies, built with the code GISSEL (Bruzual \& Charlot 1992).}
\end{figure*}

\section{Photometric characteristics} 

We investigate in this section if any correlation could e\-xist between
the photometric characteristics and the spectral classes (A, B, and C)
in the CFRS low-$z$ galaxy sample.  We do the same for the
emission-line galaxies classified within the spectral class (A).
Redshifts, $z$, apparent magnitudes, $I_{AB}$, photometric colours,
$(V-I)_{AB}$, and absolute magnitudes, $M(B_{AB})$, are given in
Table~\ref{next} for the emission-line galaxies we analysed.

\subsection{Spectral classes and the photometry}

The left panel of Figure~\ref{zI} ($I_{AB}$ {\it vs.} $z$) shows globally
that emission-line galaxies (classes A and B) have ap\-pa\-rent magnitudes
fainter than absorption-line galaxies (class~C).  Within the range
$0.2<z<0.3$, the galaxies in the class (A), (B) and (C) are
respectively within the range $20< I_{AB} < 22.5 $ mag, $19 < I_{AB}<
22.5 $ mag, and $17.5 < I_{AB} < 19.5 $ mag, approximately.
Futhermore, most galaxies spec\-tros\-co\-pi\-cal\-ly classified (A) have
$I_{AB}>20.5$ mag, whereas those classified (B) and (C) are brighter.
The apparent magnitude in this redshift range can distinguish between
the spectral classes (A) and (B)+(C).  We emphasize that class (A) is
comprised of relatively low luminosities galaxies with a noticeable
absence of luminous galaxies.  A brighter sample limited at
$I_{AB}<21$ would have a con\-si\-de\-ra\-bly smaller fraction of galaxies
(A), and would miss a large part of these strong emission-line
galaxies.  Figure~\ref{zI} (right panel) shows no distinction in the
distributions of active and H II gala\-xies; hence emission-line
galaxies in the class (A) cannot be differentiated by their apparent
magnitude. The left panel of Figure~\ref{color} ($(V-I)_{AB}$ {\it
vs.} $z$) shows that the class (A) includes mostly blue galaxies,
$(V-I)_{AB} < 0.5$, and that classes (B) and (C) includes mostly red
galaxies, $(V-I)_{AB}(z) > 0.5$.  Thus, colours can separate class (A)
and (B)+(C).  Figure~\ref{color} (right panel) shows that no
distinction in the distributions between active and H II galaxies;
hence emission-line ga\-la\-xies in class (A) cannot be differentiated by
their colour.

To summarize, classes (A) and (B)+(C) can be distinguished
photometrically in our sample.  There is the a\-na\-lo\-gous
colour-luminosity effect in Figure 5 of CFRS-VI at low redshifts, but
this relation is eliminated at the higher redshifts ($z>0.5$) by the
emergence of the population of bright blue galaxies.  Thus, we can
expect that this distinction for faint blue galaxies will also break
down at higher redshifts, if the latter are assumed to be the
descendants of the evolving bright blue population seen at $0.3<z<1$
(CFRS-VI).  Class~(A) which represents $53\%$ of all the ga\-la\-xies at
$z\leq0.3$ is mostly constituted of strong emission-line galaxies with
blue colours, $(V-I)_{AB}<0.5$, and with intrinsically faint
luminosities, $-20.5 < M(B_{AB}) < -16.5$ mag (see
Figure~\ref{histom}).  In our sample, the photometric pro\-per\-ties do
not allow distinguishing between H II and active galaxies.
\begin{figure}
\caption{\label{histom} Histogram of $B_{AB}$ absolute magnitudes of
the $138$ galaxies classified in the spectral classes (A), (B) and (C)
as defined in Section 2. Galaxies in class (A) are fainter than
$M(B_{AB})^{*}=-21$ mag (see CFRS-VI) with $\langle M(B_{AB})\rangle
=-18.5$ mag (=$L^{*}/10$).}
\end{figure} 

\subsection{The class (A) and nearby emission-line galaxies} 

We have checked whether the reddening $A_V$ could have any correlation
with the photometric properties.  In plotting $A_V$ {\it vs.}
$M(B_{AB})$, $I_{AB}$, $(V-I)_{AB}(z)$, and rest-frame $EW([O\ II])$,
no correlation is found between the different emission-line galaxies
of class (A).  This is in agreement with the results from Oey \&
Kennicutt (1993); $A_V$ does not depend on the type of galaxy from
early spirals to irregulars which should correspond to our blue
emission-line galaxy sample (see class (A) in Figure~\ref{color}).  We
calculate $\langle A_V \rangle =2.4 $ mag for spectra classified as
active galaxies, and $\langle A_V \rangle =1.5 $ for spectra
classified as H II galaxies.  This agrees with the common picture that
active galaxies are more dusty than normal galaxies (see {\it e.g.}
Cid Fernandes \& Terlevich 1995).

We have also investigated if any correlation exists bet\-ween the
rest-frame EW([O II]) and the photometric pro\-per\-ties.  
EW([O II]) is typically $10-80\ \AA$ (Figure~\ref{ew}); this confirms
that they are strong emission-line galaxies.  The range and
distribution of EWs agree with the sample of Kennicutt (1992, figure
11) of nearby galaxies of various kinds such as normal galaxies and
peculiar galaxies (not only galaxies with an active galactic nucleus,
but also those with unusual blue colours, interactions etc.). We
note that our $I$-selected sample corresponds to a $V$-selected
sample at $z\sim 0.2$, thus comparable to nearby optically-selected
galaxies.
 
We find that with the single parameter EW([O II]) one cannot
distinguish between ionization sources of different kinds, in contrast
with the diagnostic emission-line ratios. Hence, understanding the
true nature of strong emission-line spectra requires an analysis
based on several lines. In addition, the [O II] intensity depends
on the reddening, and if taken alone, one can hardly do accurate
comparisons bet\-ween galaxies of different nature.

Another interesting analysis comes from the ratio [O~II]/(H$\alpha$
+[N II]), H$\alpha$ lines are usually blended with [N~II] lines for
part of class (A) spectra.  In this ratio, lines are not corrected for
reddening or stellar absorption at H$\alpha$, similarly to Kennicutt (1992).
The median value of this ratio is $0.84$ for our emission-line
galaxies which contain [O II] within the CFRS spectral range.  The
corresponding median EW ratio is also 0.81.  The median values found
by Kennicutt (1992) are respectively 0.31 and 0.4.
Our median ratio is $0.5$ for galaxies classified as H II. 
If we account for spectra~(B) with H$\alpha$ in emission, and
sometimes [O II], our median ratio should become a bit lower, since
spectra (B) exhibit certainly lower [O II]/(H$\alpha$ + [N II]) ratios
than spectra (A). 
The difference between our sample and that of Kennicutt (1992)
arises when accounting for galaxies classified as active.  Kennicutt
finds 0.31 for several kinds of galaxies, and 0.3 excluding active
galaxy nuclei and peculiar galaxies. Although the latter sample is not
representative of the fraction of different galaxies, the large
difference may indicate some strong evolution in the spectral
properties between $z\sim 0.2$ and the nearby Universe.
\begin{figure}
\caption{\label{ew} [O II] rest-frame equivalent width {\it versus} 
 $\log ( [O\ II]\lambda3727/H\beta)$ for the galaxies in class (A) 
classified as shown in Figure 9.}
\end{figure}
 
\section{Conclusions and discussions}

From the analysis of the $138$ CFRS galaxies with $z\leq 0.3$, we have
found that: 
\begin{itemize} 
\item[(i)] $85\%$ of spectra exhibit at least a narrow-line H$\alpha$ 
in emission, and $15\%$ are purely absorption line spectra. $53\%$
exhibit several forbidden lines indicative of violent phenomena such
as enhanced star formation, winds, supernovae, or active galactic
nuclei.
\item[(ii)] The latter $53\%$ of galaxies  
show a moderate to high degree of excitation, in the range $1-2<$[O
III]/H$\beta< 10$, and have strong emission lines, with rest-frame
EW([O II]) in the range $10-80\ \AA$, rest-frame EW(H$\alpha$) in
the range $10-100\ \AA$.
\item[(iii)] $17\%$ of all galaxies at $z\leq 0.3$ have line ratios lying 
above the theoretical upper boundary of photoionization by
massive OB stars in diagnostic diagrams.  Accounting for the maximum
likely effects of stellar absorption under the Balmer lines, the 
fraction of these galaxies decreases to a minimum of $8\%$. 
Their line ratios are consistent with those observed of active galaxies. 
\item[(iv)] The $53\%$ of galaxies that have several forbidden lines are 
blue with absolute magnitude in the range $-20.5 < M(B_{AB}) < -16.5$
mag, and there is no difference in the photometric properties between
those with H II-like and active-like spectra.  Nevertheless, these
galaxies can be distinguished photometrically from the remaining
galaxies which have mostly red colours, and brighter luminosities.
\item[(v)] The median ratio [O II]/(H$\alpha$ + [N II]) ($\sim0.8$) 
is larger than that of nearby emission-line galaxies of various kinds
(Kennicutt 1992), ($\sim 0.3-0.4$), consistent with the conclusion of
more galaxies which require harder ionization sources than OB stars in
our sample.
\end{itemize} 

Our analysis shows that it is crucial to work with dia\-gnos\-tic diagrams
to understand the true nature of the emission-line galaxies in deep
surveys.  We have also determined the maximum location extent of the H
II galaxies in several diagnostic diagrams which will be useful for
further analysis (see Rola 1995).  The aim of this was to examine to
which extent OB stars could reproduce our observed line ratios, and
compare the latter with observed data.  Our results demonstrates that
at least $8\%$ of intrinsic faint blue galaxies at $z\leq0.3$ have
line ratios which require harder photoionization sources than massive
OB stars can provide.  These galaxies have line ratios consistent with
those of observed active galaxies such as Seyfert 2 or LINERs.
Further investigation is needed to determine whether the spectra of
these active galaxies are indeed being photoionised by a true active
galatic nucleus.  Finding which process can reproduce the required 
hardness is beyond the scope of this paper.

To properly estimate the evolution of the fraction of active galaxies
with epoch, large local galaxy samples, such as those studied by
Loveday {\it et al.} (1992), Marzke {\it et al.} (1994) and Metcalfe
{\it et al.}  (1991), will need to be analysed with methods similar to
what we have used in this paper to reliably establish the fraction of
active galaxies in the nearby Universe.  Until such investigations are
published, we can only compare our results to the study of Huchra \&
Burg (1992).
If our galaxies are found to contain true active galactic nuclei, 
then even the lowest allowed percentage in our sample ($8\%$) is
higher than the $2\%$ of Seyfert galaxies found by Huchra \& Burg,
which seems to indicate a significant evolution between $z\sim 0$ and
$z\sim 0.3$. However, this comparison has to be taken with caution
because of the fol\-lo\-wing reasons.  First, they do not find any faint
active galaxies ($M(B) < -19.5$) in their analysis of the CfA redshift
survey.  Second, since the definition of activity in their sample
involved cross-correlations with catalogues of known active galaxies,
it is possible that the true number of intrinsically faint active
galaxies in the local Universe have been underestimated. Third, there
is some evidence that their analysis is biased against the
identification of LINERs. Therefore, the true fraction of faint active
galaxies in the local Universe could be comparable to those found in
the CFRS survey.

In the deeper Universe, we can compare our results with the ROSAT
X-ray selected sample of Jones {\it et al.} (1995). At the lowest
X-ray fluxes they find that the survey is do\-mi\-na\-ted by a population of
narrow emission-line galaxies at typical redshifts of $z=0.2-0.4$.
Their preliminary classification with diagnostic diagrams indicates a
variety of galaxy types including Seyfert 2 and galaxies with a range
of star formation rates.  The rest-frame EW([O II]) of these galaxies
is typically $20-60\ \AA$; this is within our CFRS low-$z$ galaxy
range ($10-80\ \AA$).  Their galaxies have $M_R$ between $-20$ and
$-23$ mag, $R=16-22.5$ mag, and blue $(V-R)<0.4$, which means that our
low-$z$ sample goes fainter by about 3 ma\-gni\-tu\-des.  Therefore our
luminous $I$-selected strong emission-line galaxies may correspond to
their fainter X-ray selected emission-line galaxies. Their preliminary
study does not give the fractions of Seyfert 2 and starbursts.
However, if their galaxies and ours are similar in nature, the
contribution of soft X-ray emission in faint blue galaxies at $z\sim
0.2$ is in agreement with our observations of line ratios consistent
with active galaxies. The X-ray emission is also observed in their
sample in some H II galaxies (or starbursts).  Ac\-tual\-ly some of our
spectra classified as either H II galaxies or active galaxies
respectively to their line ratios may have the same origin of ionizing
sources harder than OB stars. 
 
Our results can be related to those of $B$-selected samples (see {\it
e.g.} Ellis {\it et al.} 1995) whose redshifts are mainly below $0.5$.
Our bluest galaxies, which correspond roughly to the $53\%$ of
galaxies with several strong emission lines, will be present in these
samples, and at least $20\%$ of these ga\-la\-xies have emission-line
ratios which require harder ionization sources than OB stars.  Our
analysis provides evidence that harder ionization processes contribute
to the rapidly e\-vol\-ving intrinsic faint blue galaxy population.
Therefore, understanding the nature of these galaxies with active-like
spectra (which may well not have true active nuclei) is likely be
required before we fully understand the evolution of this po\-pu\-la\-tion.
Futhermore, the CFRS luminosity function (CFRS-VI) shows that the
bright blue population at $z>0.5$ could be the parent of the faint
blue galaxies at low redshifts.
It will be important, but observationally
challenging, to extend this diagnostic analysis to higher redshifts,
$0.3 < z < 1$, where the evolution of the luminosity function of blue
ga\-la\-xies continues to be strongly evident.

\section*{Acknowledgments} 

We gratefully acknowledge Elena and Roberto Terlevich, and Catherine
Boisson for useful discussions and helpful suggestions. We thank
Hector Flores for his help. LT thanks Richard Ellis for financial
support at the Institute of As\-tro\-no\-my. CR acknowledges the support of
the JNICT (Portugal) by the BD/3811/94 PRAXIS XXI grant.

\appendix 
 
\bsp 
 

\begin{thebibliography}{99} 
\bibitem{b1} Anders E., Grevesse  N. ,1989, {\it Geochimica \& 
Cosmochimica Acta}, {\bf 53},  197. 
\bibitem{b1} Baldwin, J. A., Phillips, M. M., Terlevich, R., 1981, {\it PASP}, 
 {\bf 93}, 5 (BPT). 
\bibitem{b1} Binette, L., Magris, C. G., Stasi\'nska, G., Bruzual, A. G., 
1994, {\it A\& A}, {\bf 292}, 13.
\bibitem{b1} Boroson, T. A., Salzer, J. J., Trotter, A., 1993, {\it ApJ}, 
{\bf 412}, 524.
\bibitem{b1} Broadhurst, T. J., Ellis, R. S., Shanks, T., 1988, {\it MNRAS}, 
 {\bf 235}, 827. 
\bibitem{b1} Bruzual, G., Charlot, S. 1993, {\it ApJ}, {\bf 405}, 538. 
\bibitem{b1} Cid Fernandes, R., Terlevich, R., 1995, {\it MNRAS}, 
{\bf 272}, 423.
\bibitem{b1} Colless, M. M., Ellis R. S., Taylor, K., Hook, R. N., 1990, 
 {\it MNRAS}, {\bf 244}, 408.
\bibitem{b1} Crampton, D., Le F\`evre, O., Lilly, S. J., Hammer, F., 
1995a, {\it ApJ}, {\bf 455}, 96 (CFRS-V). 
\bibitem{b1} Dinerstein, H. L., Shields, G. A., 1986, {\it ApJ}, {\bf 311}, 45.
\bibitem{b1} Ellis, R. S., Colless, M., Broadhurst, T., Heyl, J., 
Glazebrook, K., 1995, {\it MNRAS}, preprint.   
\bibitem{b1} Garnett, D. R., 1989, {\it ApJ}, {\bf 345}, 282. 
\bibitem{b1} Glazebrook, K., Ellis, R., Colless, M., Broadhurst, T.,
Allington-Smith, J., Tanvir, N., 1995, {\it MNRAS}, {\bf 273}, 157. 
\bibitem{b1} Hammer, F., Crampton, D., Le F\`evre, O., Lilly, S., 1995a, 
{\it ApJ}, {\bf 455}, 88 (CFRS-IV).
\bibitem{b1} Jones, L. R. {\it al.}, 1995 
  in {\it Wide field spectroscopy and  the  distant  universe}, 
  35th Hertsmonceux Conference, 
  eds. S. Maddox et A. Aragon-Salamanca, World Scientific, p. 339. 
\bibitem{b1} Huchra, J., Burg, R., 1992, {\it ApJ}, {\bf 393}, 90.  
\bibitem{b1} Kennicutt, R. C., Edgar, B. K., Hodge, P. W., 
1989, {\it ApJ}, {\bf 337}, 761. 
\bibitem{b1} Kennicutt, R. C., 1992, {\it ApJ}, {\bf 388}, 310. 
\bibitem{b1} Kurucz, R., 1992, in {\it The Stellar Population of galaxies},
eds. Barbuy \& Renzini, Kluwer Academic Publishers, 225.  
\bibitem{b1} Le F\`evre, O., Crampton, D., Lilly, S. J., Hammer, F., 
 Tresse, L., 1995a, {\it ApJ}, {\bf 455}, 60 (CFRS-II).  
\bibitem{b1} Lilly, S. J., Hammer, F., Le F\`evre, O., Crampton, D., 
 1995b, {\it ApJ}, {\bf 455}, 75 (CFRS-III). 
\bibitem{b1} Lilly, S. J., Le F\`evre, O., Crampton, D., Hammer, F., 
 Tresse, L., 1995a, {\it ApJ}, {\bf 455}, 50 (CFRS-I). 
\bibitem{b1} Lilly, S. J., Tresse, L., Hammer, F., Crampton, D., 
 Le F\`evre, O., 1995c, {\it ApJ}, {\bf 455}, 108 (CFRS-VI). 
\bibitem{b1} Loveday, J., Peterson, B. A., Efstathiou, G., Maddox, S. J., 
1992, {\it ApJ}, {\bf 390}, 338. 
\bibitem{b1} Maeder, A., 1990, {\it A\& AS}, {\bf 84}, 139.
\bibitem{b1} Marzke, R. O., Geller, M. J., Huchra, J. P, Corwin, H., 1994, 
{\it AJ}, in press.   
\bibitem{b1} Mathis, J. S., 1986, {\it PASP}, {\bf 98}, 995. 
\bibitem{b1} McCall, M. L., Rybski, P. M., Shields, G. A. 1985, {\it ApJS}, 
{\bf 57}, 1. 
\bibitem{b1} McGaugh, S., 1991, {\it ApJ}, {\bf 380}, 140. 
\bibitem{b1} Metcalfe, N., Fong, R., Shanks, T., Kilkenney, D., 1989, 
{\it MNRAS}, {\bf 237}, 207.
\bibitem{b1} Oey, M. S. and Kennicutt, R. C., 1993, {\it ApJ}, {\bf 411}, 137.
\bibitem{b1} Osterbrock, D. E., 1989, {\it Astro. of Gaseous Nebulae \& 
 Active Galactic Nucleii}, Univ. Sci. Books.
\bibitem{b1} Rola, C., 1995, Ph.D. thesis, University of Paris VII, France. 
\bibitem{b1} Rola, C., Pelat, D., 1994, {\it A\& A}, {\bf 287}, 676. 
\bibitem{b1} Salzer, J. J., MacAlpine, G. M., Boroson, T. A., 1989, 
{\it AJSS}, {\bf 70}, 479.
\bibitem{b1} Seaton, M. J., 1979, {\it MNRAS}, {\bf 187}, 73.
\bibitem{b1} Shields, J. C., Kennicutt, R. C., 1996, {\it ApJ}, preprint.
\bibitem{b1} Stasi\'nska, G., 1990, {\it A\& AS}, {\bf 83}, 501. 
\bibitem{b1} Stasi\'nska, G., Leitherer, K., 1996, {\it ApJS}, preprint.
\bibitem{b1} Terlevich, R., Tenario-Tagle, G., Franco, J., Melnick, J., 
Boyle, B. J., 1993, in {\it The nearest active galaxies},
eds. J. Beckman, L. Colina, H. Netzer, Nuevas Tendencias, p. 181.
\bibitem{b1} Tresse, L., Hammer, F., Le F\`evre, O., Proust, D., 1993, 
{\it ApJ}, {\bf 277}, 53.
\bibitem{b1} Tresse, L., Rola, C., Hammer, F., Stasi\'nska, G., 1995a, 
  in {\it Wide field spectroscopy and  the  distant  universe}, 
  35th Hertsmonceux Conference, 
  eds. S. Maddox et A. Aragon-Salamanca, World Scientific, p. 290. 
\bibitem{b1} Veilleux, S., Osterbrock, D. E., 1987, {\it ApJS}, {\bf 63}, 
 295 (VO).
\end{thebibliography}
\end{document}